\documentclass{book}
\usepackage[dvips]{epsfig}
\usepackage{epsf}
\begin{document}
\begin{titlepage}
\vspace{2cm}
\hfill{\Large{ SLAC-PUB 7383}}

\hfill{\Large{ hep-ph/9612348}}

\hfill{\large{ December 1996}}

\vspace{0.8cm}
\centerline{\LARGE{\bf Exact Results on ${\cal O}(\alpha^2)$ Single Bremsstrahlung}} 
\vspace{0.4cm}
\centerline{\LARGE{\bf Corrections to Low Angle Bhabha Scattering}}
\vspace{0.4cm}
\centerline{\LARGE{\bf at LEP/SLC Energies}}

\vspace{0.8cm}
\centerline{\LARGE{\bf Michael Melles}}
\vspace{0.2cm}
\centerline{\large{\it Stanford Linear Accelerator Center}}
\vspace{0.2cm}
\centerline{\large{\it Stanford University, Stanford, CA 94309}}

\section*{Abstract}
\noindent In this thesis exact results on ${\cal O}(\alpha^2)$ single bremsstrahlung
corrections to low angle Bhabha scattering at LEP/SLC energies are given.
The calculation represents the last outstanding theoretical second order
subleading electroweak contribution for that process, needed to determine the 
experimental luminosity at second generation LEP detectors below the $0.1\%$
precision threshold. The exact, fully differential result is 
obtained by employing analytical
as well as computer-algebraic methods and includes terms up to ${\cal O}(0.05\%)$ 
relative to the Born cross section. The initial output of over $20,000$ terms
could be reduced to $90$, only $18$ of which are shown to be numerically
relevant and for which a simple logarithmic ansatz is derived, that is in
remarkable agreement with the complete answer.  
Strong consistency checks are performed,
including Ward-Takahashi identities and tests on the right infrared limit
according to the Yennie, Frautschi and Suura program. Monte Carlo results 
for the integrated cross section are compared with existing
calculations in the leading logarithmic approximation for a chosen set of 
experimental cuts. The size of the missing subleading 
terms is found to
be small but non negligible in the context of setting stringent limits on
Standard Model predictions and thus its realm of validity. 

\vspace{0.8cm}
\centerline{ To be published in Acta Physica Polonica}

\vspace{0.8cm}
\noindent Research supported by the Department of Energy under grants
DE-AC03-76SF00515, DE-FG05-91ER40627 and by the Deutsche Forschungsgemeinschaft
(DFG)
\end{titlepage}

\tableofcontents

\part{${\cal O}(\alpha)$ Corrections to $e^-\;+\;e^+ \longrightarrow e^-\;+\;e^+ \; +\; \gamma$}

\chapter{Introduction} \label{sec:intro}

\noindent High precision measurements at the LEP/SLC colliders have made 
tremendous improvements over the last few years. At LEP, for instance, first
generation detectors have measured the absolute luminosity with an accuracy of
$0.3-0.5\%$ \cite{Piet94}. Present detectors have 
reached a precision below the $1pm$ threshold \cite{Piet94}. This level of precision
opens up a wide range of new stringent constraints that can be specified not
only for physics beyond the Standard Model, but also for the Higgs sector 
and other parameters of the electroweak theory \cite{altar95}. While LEP $I\!I$ is
now on line to probe the $W^+$, $W^-$ aspect of the
GSW-model, there is still a lot of interest in the LEP $I$ energy regime of 
the intermediary neutral vector boson.

\noindent In general, the $Z$ line shape is described by three parameters: $M_Z$, 
$\Gamma_Z$ and $\sigma^0$ \cite{Piet94}. The luminosity normalizes the line shape 
cross section and it thus directly influences the accuracy of the $\sigma^0$
measurement. In order to determine the luminosity $L$, one needs to count the
number of events $N$ for a process in which the cross section $\sigma$
is "known":

\begin{equation}
L\;\;=\;\; \frac{N}{\sigma}
\end{equation}

\noindent The "known" process at LEP/SLC is low angle Bhabha scattering 
\cite{Bhabha35}. More than $99\%$ of this cross section is due to t-channel photon 
exchange \cite{Been92} and can in principle be calculated in QED with arbitrary 
precision. The remaining part of the cross section is due to
s-channel $Z$, t-channel 
$\gamma$ interference, and its relative contribution decreases with 
decreasing scattering angle \cite{Piet94}.

\noindent The overall accuracy of the "known" cross section is very 
important for high precision measurements of Standard Model parameters
since a relative luminosity error $\frac{\delta L}{L} \; = \; 10^{-3}$,
for example, changes $\sigma_0$ by $42pb$ and $N_{\nu}$, the number of 
massless neutrino generations, by $0.0075$ \cite{Piet94}. Also $\Gamma_e \; \equiv
\; \Gamma(Z \rightarrow e^+\!+e^-)$ is directly affected by the precision 
of $L$ since \cite{Jadtalk1}

\begin{equation}
\Gamma_e\!^2\;\;=\;\; \frac{M_Z\!^2 \sigma_e\!^0{\Gamma\!_Z}\!^2}{12\pi} \;,
\end{equation}

\noindent which is one of the main sources of our knowledge of the electroweak
mixing angle \cite{Jadtalk1}. It is is thus necessary to have theoretical 
calculations that can
match the experimental accuracy below the $1pm$ mark. As can be seen from Refs.
\cite{JadWas95,Gatlin94,Yellow95}, this has not yet been achieved, 
although initial
state s-channel ${\cal O}(\alpha^2)$ corrections, including two loop effects have
been calculated \cite{berends87}. The result of Ref. \cite{berends87} 
for one virtual and one
hard photon, however, is not suited for a Monte Carlo implementation since  
it was calculated in an "inclusive" way, i.e. it is not differential in the
photon angles, which prohibits any adaptation to a given detector geometry.
This is reflected in the form that the 
aspect of the theoretical uncertainty that
contributes most to the present $1.6pm$ precision level is 
given by the missing parts of
the second order subleading terms originating from the one photon
bremsstrahlung process \cite{JadWas95}. 
In this thesis, the exact ${\cal O}(\alpha)$ virtual corrections to the
single hard bremsstrahlung cross section at $\sqrt{s}=M_Z$ are
calculated up to the precision
required in the above context, i.e. up to ${\cal O}(\leq 0.05\%)$.  
This means, for instance, that two photon exchange contributions can be 
neglected \cite{JadRiWa92} as well as t-channel $Z$ exchange diagrams \cite{JadWas95}.   
The result is calculated in differential form and Monte Carlo \cite{sobol}
results then 
give the integrated cross section that is relevant for the 
theoretical prediction of the low angle Bhabha scattering cross section, 
needed to determine the luminosity $L$.


\noindent This work is structured as follows:

\noindent {\bf Chapter \ref{sec:defid}} describes the conventions and defines 
the notation used in this thesis. A brief section is devoted to scalar 
integrals and how to approach complex loop integrals in the context of using 
an algebraic manipulation language. It also gives an expression for the 
single bremsstrahlung tree level amplitude $A^{TL}\!\!\!\!\!_{e^-}$, which
will prove to be pivotal for the final form of the exact differential result
given in chapter \ref{sec:exres}.

\noindent In {\bf chapter \ref{sec:singBramp}} the details of 
the calculation of the ${\cal O}(\alpha)$ virtual corrections to the single hard 
bremsstrahlung cross section are presented. The nature of the problem of 
evaluating higher order Feynman amplitudes brings with it a somewhat 
unavoidable technical aspect in the various sections. This chapter contains 
the new results presented in this work in conjunction with the appendices.

\noindent A summary of the various expressions of chapter \ref{sec:singBramp}
is given in {\bf chapter \ref{sec:exres}} as the complete differential result.
First numerical results are presented and discussed.

\noindent {\bf Chapters \ref{sec:YFS} and \ref{sec:gaugevar}} 
are devoted to demonstrating that the presented 
exact differential result satisfies various internal consistency checks and
has all the properties expected from general theoretical arguments.

\noindent The total cross section resulting from the differential expression
in chapter \ref{sec:exres} is obtained by means of a Monte Carlo 
phase space integration and numerical comparisons with other published 
leading log (LL) type calculations are presented in {\bf chapter 
\ref{sec:MCres}} for cuts similar to those used at the luminosity
detector SICAL at ALEPH in LEP \cite{JadRiWa95}. Also, a new approximation is 
derived that shows excellent agreement with the exact result and the overall
size of the subleading terms is discussed.

\noindent Finally, {\bf chapter \ref{sec:concl}} 
contains a summary of the results obtained in
this thesis in the context of the present status of high precision
radiative corrections and also concluding remarks. 


\chapter{Basic Definitions and Identities} \label{sec:defid}

\noindent In this chapter the basic notations and definitions will be discussed as well as some identities that follow from the high energy limit. If nothing else is stated explicitly, the conventions of Bjorken and Drell \cite{bjdrell1} are used throughout this work. Since the overall objective of this thesis
is a high precision calculation of ${\cal O}(\alpha^2)$ radiative corrections 
to a high energy process, it will prove to be very convenient to make use of the so called high energy limit, where terms of order $\frac{{m_e}^2}{-t}$ can be neglected. As a consequence of rendering the electrons (positrons) massless, their helicity will be conserved. This then leads to the very useful concept of introducing fermion helicity states \cite{causm81} which are defined here by the following Dirac notation \cite{gastwu}:

\begin{eqnarray}
|P,\lambda \rangle \,\, \equiv \,\, u_{\lambda}\!(P) \,\, = \,\, v_{-\lambda}\!(P) \label{eq:hel1} \\
\langle P,\lambda | \,\, \equiv \,\, \bar{u}_{\lambda}\!(P) \,\, = \,\, \bar{v}_{-\lambda}\!(P) \label{eq:hel2} 
\end{eqnarray}

\noindent Eqs. \ref{eq:hel1}, \ref{eq:hel2} hold for arbitrary four-momentum $P$ with $P^2=0$ and $u,v$ defined as in Ref. \cite{bjdrell1}. The normalization is fixed by demanding that

\begin{equation}
\langle P,\lambda | \, \gamma_{\mu} \, |P,\lambda \rangle \,\, = \,\, 2P_{\mu} \label{eq:norm1}
\end{equation}

\noindent For two arbitrary massless four-momenta, $P$ and $Q$, the following equations hold:

\begin{eqnarray}
\langle P,\lambda |Q,\lambda \rangle \, = \, 0 \, ; \, \langle P,-\lambda |P,\lambda \rangle \, = \, 0 \, ; \, \rlap/ \!P \, |P,\lambda \rangle \, = \, 0 \label{eq:prop1} \\
\langle P,-\lambda |Q,\lambda \rangle \, = \, - \, \langle Q,-\lambda |P,\lambda \rangle \, ; \, \rlap/ \!P \, = \, |P,\lambda \rangle \langle P,\lambda | + |P,-\lambda \rangle \langle P,-\lambda | \label{eq:prop2} \\
\langle P,-\lambda |Q,\lambda \rangle \, \langle Q,\lambda |P,-\lambda \rangle \,\, = \,\, 2PQ \,\, = \,\,(P+Q)^2 \label{eq:prop3}
\end{eqnarray}

\noindent Furthermore, for arbitrary four-momenta $k_i$:

\begin{eqnarray}
\langle P,\lambda | \, \rlap/ k_1 \, . . . \, \rlap/ k_n \, |Q,\lambda \rangle &=& \langle Q,-\lambda | \, \rlap/ k_n \, . . . \, \rlap/ k_1 \, |P,- \lambda \rangle \,\,\,\,\,\,\,\,\, , n \,\,\,\, odd \label{eq:nodd} \\ 
\langle P,-\lambda | \, \rlap/ k_1 \, . . . \, \rlap/ k_n \, |Q,\lambda \rangle &=& - \, \langle Q,-\lambda | \, \rlap/ k_n \, . . . \, \rlap/ k_1 \, |P,\lambda \rangle \;\;\;\;\; , n \,\, even \label{eq:neven}
\end{eqnarray}

\noindent An important identity that will be used extensively is the Fierz-identity \cite{itzzub} and holds for arbitrary spinors $|A,\lambda \rangle \, , \, |B,\lambda' \rangle \, , \, |C,\mu \rangle $ and $ |D,\mu' \rangle $ :

\begin{equation}
\langle A,\lambda | \, \gamma_{\nu} \, |B,\lambda \rangle \, \langle C,\mu | \, \gamma^{\nu} \, | D, \mu \rangle \,\, = \,\, 2 \, \langle A, \lambda | X,-\lambda \rangle \, \langle Y,-\lambda | B, \lambda \rangle \;, \label{eq:Fierz}
\end{equation}

\noindent where

\begin{equation}
(X,Y) \,\,\,=\,\,\, \left\{ \begin{array}{ll} (C,D) & \mbox{$, \lambda=\mu$} \\
 (D,C) & \mbox{$, \lambda=-\mu$} \end{array} \right. \\ \label{eq:Fierzparam}
\end{equation}

\noindent It is sometimes convenient to simply write 

\begin{equation}
\!\!\!\!\!\!\!\!\!\!\!\!\!\!\!\!\!\!\!\!\!\!\!\! {{\langle P,Q \rangle}_{\lambda} \,\,\, \equiv \,\,\,  \langle P,-\lambda | Q, \lambda \rangle} \label{abbrev1}
\end{equation}

\noindent The "magic" polarization vector of Xu, Zhang and Chang is given by \cite{XZC87} :

\begin{equation}
\!\!\!\!{\varepsilon_{\mu} \! (K,h,\rho) \,\,\, = \,\,\, \frac{\rho}{\sqrt{2}} \, \frac{\langle h, -\rho | \, \gamma_{\mu} \, | K , -\rho \rangle }{ {\langle h, K \rangle}_{\rho}}} \; , \label{eq:epsmu}
\end{equation}

\noindent where $K$ is the photon four-momentum, $\rho$ its polarization and $h$ a reference momentum with $h^2=0$. Choosing $h$ amounts to a gauge choice and this freedom can be exploited to greatly simplify calculations with bremsstrahlung effects \cite{gastwu}. If the incoming electron has a four-momentum $P$, the outgoing one a four-momentum $Q$ and if its helicity is denoted by $\lambda$, the "magic" choice in the present calculation turns out to be :

\begin{equation}
h \,\,\,=\,\,\, \left\{ \begin{array}{ll} P & \mbox{$, \lambda=\rho$} \\
Q & \mbox{$, \lambda=-\rho$} \end{array} \right. \\ \label{eq:hdef}
\end{equation}

\noindent For completeness, there are three other helicity dependent four-momenta used in this work, $h', \widehat{h}$ and $ \widehat{h'}$, where analogously for the positron line

\begin{equation}
h' \,\,\,=\,\,\, \left\{ \begin{array}{ll} P' & \mbox{$, \lambda'=\rho$} \\
Q' & \mbox{$, \lambda'=-\rho$} \end{array} \right. \\ \label{eq:h'def}
\end{equation}

\noindent $\widehat{h}$ and $ \widehat{h'}$ denote the helicity flipped choices of $h$ and $h'$ respectively.

\noindent It is also helpful to give $\rlap/ \varepsilon$ explicitly, since it is the object appearing in expressions for Feynman amplitudes :

\begin{equation}
\rlap/ \varepsilon \,\,\, = \,\,\, \frac{\rho \sqrt{2}}{{\langle h,K \rangle}_{\rho}} \, ( \, | K , -\rho \rangle \langle h, -\rho | \, + \, | h , \rho \rangle \langle K, \rho | \, ) \label{eq:eps}
\end{equation}

\noindent As an application, the "magic" properties of $\varepsilon_{\mu}$ are used in proving the following important identity for $\lambda = \rho$:

\begin{eqnarray}
0 &=& PQ \, \langle Q',\lambda'| \, \rlap/ \varepsilon \, |P',\lambda' \rangle 
\, \langle Q,\lambda| \, \rlap/ \!K \, |P,\lambda  \rangle \,\, - \nonumber \\
& & Q \varepsilon \,\,\, \langle Q',\lambda'| \, \rlap/ \!P \, |P',\lambda' \rangle \, \langle Q,\lambda| \, \rlap/ \!K \, |P,\lambda  \rangle \,\, + \nonumber \\
& & PK \, Q \varepsilon \, \langle Q',\lambda'|\, \gamma_{\mu} \, |P',\lambda' \rangle \, \langle Q,\lambda| \, \gamma^{\mu} \, |P,\lambda \rangle \label{eq:id1}
\end{eqnarray}

\noindent Eq. \ref{eq:id1} can be proved by observing that  

\begin{eqnarray}
PK \, \langle Q,\lambda| \, \gamma^{\mu} \, |P,\lambda \rangle &=& P^{\mu} \, \langle Q,\lambda| \, \rlap/ \!K \, |P,\lambda  \rangle \, - \, \frac{1}{2} \, \langle Q,\lambda| \, \rlap/ \!P \, \gamma^{\mu} \, \rlap/ \!K \, |P,\lambda  \rangle \\
PQ \, \langle Q,\lambda| \, \rlap/ \!K \, |P,\lambda  \rangle &=& \frac{1}{2} \, \langle Q,\lambda| \, \rlap/ \!P \, \rlap/ \!Q \, \rlap/ \!K \, |P,\lambda \rangle 
\end{eqnarray}

\noindent Thus, it remains to be shown that

\begin{eqnarray}
0 &=& \langle Q',\lambda'| \, \rlap/ \varepsilon \, |P',\lambda' \rangle \, \langle Q,\lambda| \, \rlap/ \!P \, \rlap/ \!Q \, \rlap/ \!K \, |P,\lambda \rangle\,\, - \nonumber \\ & & Q \varepsilon \, \langle Q',\lambda'|\, \gamma_{\mu} \, |P',\lambda' \rangle \, \langle Q,\lambda| \, \rlap/ \!P \, \gamma^{\mu} \, \rlap/ \!K \, |P,\lambda \rangle \;, 
\end{eqnarray}

\noindent which follows from the transversality condition $\varepsilon K = 0$ and the "magic" properties

\begin{equation}
\!\!\!\!\!\!\!\!\!\!\!\!\!\!\!\!\!\!\!\! {\rlap/ \varepsilon \, |P,\lambda \rangle \, = \, 0 \,\,\,\, \& \,\,\,\, \{ \, \rlap/ \!P , \rlap/ \varepsilon \, \} \, = \, 0 \,\,\,\,\,\,\,\,\,\,\,\, , \lambda = \rho}  
\end{equation}

\noindent An analogous identity holds for $\lambda = - \rho$ and reads:

\begin{eqnarray}
0 &=& PQ \, \langle Q',\lambda'| \, \rlap/ \varepsilon \, |P',\lambda' \rangle
\, \langle Q,\lambda| \, \rlap/ \!K \, |P,\lambda  \rangle \,\, - \nonumber \\
& & P \varepsilon \,\,\, \langle Q',\lambda'| \, \rlap/ \!Q \, |P',\lambda'    \rangle
\, \langle Q,\lambda| \, \rlap/ \!K \, |P,\lambda  \rangle \,\, + \nonumber \\
& & QK \, P \varepsilon \, \langle Q',\lambda'|\, \gamma_{\mu} \, |P',\lambda'
\rangle \, \langle Q,\lambda| \, \gamma^{\mu} \, |P,\lambda \rangle            \label{eq:id2}
\end{eqnarray}

\noindent Eq. \ref{eq:id2} can be directly obtained from Eq. \ref{eq:id1} through the crossing rule:

\begin{equation}
\!\!\!\!\!\!\!\!\!\!\!\!\!\!\!\!\!\!\!\!\!\!\!\!\!\! {P \, \leftrightarrow \, - Q \,\,\, , \,\,\, \lambda \, \rightarrow \, - \lambda \label{eq:cross1}}
\end{equation}

\noindent The usefulness of the helicity dependent spinor notations \ref{eq:hdef}, \ref{eq:h'def} can be exemplified by stating the identities \ref{eq:id1}, \ref{eq:id2} through a helicity independent notation:

\begin{eqnarray}
0 &=& h \widehat{h} \, \langle Q',\lambda'| \, \rlap/ \varepsilon \, |P',\lambda' \rangle
\, \langle Q,\lambda| \, \rlap/ \!K \, |P,\lambda  \rangle \,\, - \nonumber \\
& & \widehat{h} \varepsilon \,\,\, \langle Q',\lambda'| \, \rlap/ h \, |P',\lambda'    \rangle
\, \langle Q,\lambda| \, \rlap/ \!K \, |P,\lambda  \rangle \,\, + \nonumber \\
& & hK \, \widehat{h} \varepsilon \, \langle Q',\lambda'|\, \gamma_{\mu} \, |P',\lambda'
\rangle \, \langle Q,\lambda| \, \gamma^{\mu} \, |P,\lambda \rangle            \label{eq:id3}
\end{eqnarray}

\noindent Note in particular that the polarization vector $\varepsilon_{\mu}$ in \ref{eq:epsmu} is invariant under the transformation \ref{eq:cross1}, which relates initial to final state electron-line radiation and thus insures that the same polarization vector be used for one line in each helicity case.  

\noindent The following definitions will also be extensively used throughout this work, where the four-momenta are those of \ref{eq:hdef} and \ref{eq:h'def}:

\begin{eqnarray}
\alpha & \equiv & \;\; 2PK \,\,\,\;\; = \,\, -(P-K)^2  \label{eq:aldef} \\
\beta & \equiv & \;\; 2QK \,\,\,\;\; = \;\;\,\, (Q+K)^2 \label{eq:bedef} \\
s \, & \equiv & \;\; 2PP' \,\,\,\; = \;\;\;\, (P+P')^2 \label{eq:sdef} \\
s' & \equiv & \;\; 2QQ' \,\,\,\; = \;\;\;\, (Q+Q')^2 \label{eq:s'def} \\
t \, & \equiv & -2PQ \,\,\;\; = \,\,\,\;\; (P-Q)^2 \label{eq:tdef} \\
t' & \equiv & -2P'Q' \,\, = \,\,\,\, (P'-Q')^2 \label{eq:t'def} 
\end{eqnarray}

\noindent It should be mentioned that the fine structure constant $\alpha = \frac{e^2}{4 \pi} = \frac{1}{137}$ will not appear explicitly in equations in this work unless specifically mentioned in order to avoid confusing notation. 

\noindent The t-channel process, as shown in Fig. \ref{fig:Atrlvl}, also 
allows for $Z$-boson exchange in the present calculation, mainly to be able 
to incorporate s-channel $Z$, t-channel $\gamma$ interference and, 
at a later stage,
to possibly allow for wide-angle applications. The propagator $G_{\lambda,\lambda\!'}(t')$ is then given by \cite{Kleiss90}:

\begin{equation}
G_{\lambda,\lambda\!'}(t') \;\; = \;\; \frac{1}{t'} \; + \; \frac{[(1-\lambda) - 4 sin^2 \theta_w][(1-\lambda')- 4 sin^2 \theta_w]}{4 sin^2 2\theta_w (t' - {M\!_Z}^2 )} \;, \label{eq:prop}
\end{equation}

\noindent where $\theta_w$ denotes the weak mixing angle and ${M\!_Z}$ the mass of the intermediary $Z$-boson \cite{Steinb91}. In the s-channel a correction coming from the imaginary part of the self-energy correction \cite{Kleiss90} needs to be taken into account and leads to the replacement rule ${M\!_Z}^2 \longrightarrow {M\!_Z}^2 - i s \frac{\Gamma\!_Z}{M\!_Z}$, where $\Gamma\!_Z$ is the width of the $Z$-boson.

\section{Scalar Integrals} \label{sec:scalintdef}

\noindent All results that are given below involving loop integrals are expressed in terms of so called scalar integrals, i.e. four-dimensional integrals over Minkowski space with no tensor-structure in the numerators. The analytic solutions of integrals of this type have long been known in the literature \cite{thooft79} and recently, a new way of numerically implementing these results in a more stable way has been proposed \cite{verma90}. The results of Ref. \cite{verma90} are those that have been used in this work to calculate scalar integrals by means of a numerical package called $f\!f$, as well as their reduction scheme, which decomposes tensor integrals into a linear combination of the easier scalar integrals. Wherever possible, the package results have been compared to known analytic results in certain 
limits \cite{thooft79,dittm93} and have shown very good agreement. The reduction scheme, implemented in the algebraic computer-language FORM \cite{verma90}, has also passed all tests that were applied to it with relatively simple test-cases. It is essential for a numerical analysis that the algorithms, especially for the more complicated scalar integrals with three and four denominators, are reliable and indicate situations where the required cancelations become too involved to be handled within a given level of accuracy. This is one of the most important advantages the $f\!f$-package possesses, where roughly $80\%$ of the programmed code ($\approx 50000$ lines) is devoted to just that 
\cite{oldenb}. 
The scalar integrals used in the present calculation are defined as follows:

\begin{eqnarray}
{X^k}\!\!\!\!_0  & \equiv &  \frac{1}{i\pi^2} \int \!\! \frac{\mu^{\epsilon} \;d^n l}{(l^2 - {m_1}\!^2) ((l+p_1)^2 - {m_2}\!^2) ... ((l+p_1+...+p_{k-1})^2 - {m_k}\!^2)+i\epsilon} \nonumber \\ && \label{eq:scalardef}
\end{eqnarray}
 
\noindent Throughout this work only scalar integrals up to $k=4$ will be necessary and are going to be denoted by 

\begin{eqnarray}
{X^1}\!\!\!\!_0 & \equiv & A_0 (m_1) \label{eq:A0def} \\
{X^2}\!\!\!\!_0 & \equiv & B_0 ({p_1}\!^2,m_1,m_2)  
\label{eq:B0def} \\
{X^3}\!\!\!\!_0 & \equiv & C_0 ({p_1}\!^2,{p_2}\!^2,(p_1+p_2)^2,m_1,m_2,m_3) \label{eq:C0def} \\
{X^4}\!\!\!\!_0 & \equiv & D_0 ({p_1}\!^2,{p_2}\!^2,{p_3}\!^2,{p_4}\!^2,(p_3+p_4)^2,(p_1+p_4)^2,m_1,m_2,m_3,m_4) \label{eq:D0def} 
\end{eqnarray}

\noindent It should be noted that, as suggested by the notation in Eq. 
\ref{eq:scalardef}, the scalar integrals \ref{eq:A0def}, \ref{eq:B0def} are 
defined through the n-dimensional regularization method \cite{ramond,ryder}. 
The choice of the scale $\mu$ should not affect any result.
In order to write Eq. \ref{eq:scalardef} in an easy to recognize manner for arbitrary $k$, all momenta $p_i$ are assumed to be incoming in a Feynman-diagram. For the calculation of ${\cal O}(\alpha^2)$ single bremsstrahlung corrections in this work, however, all physically outgoing particles will also be denoted by outgoing four-momenta, such that the scalar product of incoming with outgoing
momenta will always be positive. Since massless spinors are used in this 
thesis, all terms homogeneous in $m_e$ will be dropped out of the expressions
derived below. The masses, however, will be kept as regulators of "unphysical"
collinear divergences inside the scalar integrals. 

\section{Complex Loop Integrals and Bilinear Covariants}

\noindent When using algebraic manipulation languages to calculate higher 
order Feynman diagrams, it is important to reduce the sometimes very involved
matrix expressions in numerators of the resulting integrals. Since these
terms will eventually be evaluated between spinors, it is desirable to 
reduce strings of Dirac gamma matrices $-$ up to seven in integrals that will 
be treated in chapter \ref{sec:singBramp} $-$ to terms multiplying bilinear 
covariants \cite{bjdrell1}. The transformation behaviour of the various 
basis terms is that of a scalar ($1$), a pseudoscalar ($\gamma_5$), a vector
($\gamma_{\mu}$), an axial-vector ($\gamma_5\gamma_{\mu}$) and that of a 
second rank tensor ($\sigma_{\mu \nu} \equiv \frac{i}{2} 
[\gamma_{\mu},\gamma_{\nu}]$).
Expanding an arbitrary string of slashed four-vectors, $G \equiv \not 
\!a_1... \not \!a_n$, yields: 

\begin{eqnarray}
G &=& \frac{1}{4} Tr(G) 1 + \frac{1}{4} Tr(\gamma_5 G) \gamma_5 + \frac{1}{4}
Tr(\gamma_{\mu}G) \gamma^{\mu} \nonumber \\
&& - \frac{1}{4} Tr(\gamma_5\gamma_{\mu}G) \gamma_5\gamma^{\mu} + \frac{1}{8}
Tr(\sigma_{\mu \nu}G) \sigma^{\mu \nu} \label{eq:bilindecomp}
\end{eqnarray}

\noindent  
It can be easily checked that
if $G$ is one of the stated basis vectors, Eq. \ref{eq:bilindecomp} holds
in each case. The use of $\gamma_5$ can be avoided in n-dimensions, but
the algebra in the end always assumes that the spinor space remains 
four-dimensional.
The reduction \ref{eq:bilindecomp} also allows the application of the 
entire spinology that was described in the beginning of this chapter, since
the terms in \ref{eq:bilindecomp} are easily evaluated between spinors.
Ref. \cite{thesis} contains a program written in FORM that produces
the presented reduction in n-dimensions and is employed in most 
of the other routines used in this thesis.

\section{The Tree Level Single Bremsstrahlung Amplitudes} \label{sec:Atreel}

\begin{center}
\begin{figure}
\centering
 \epsfig{file=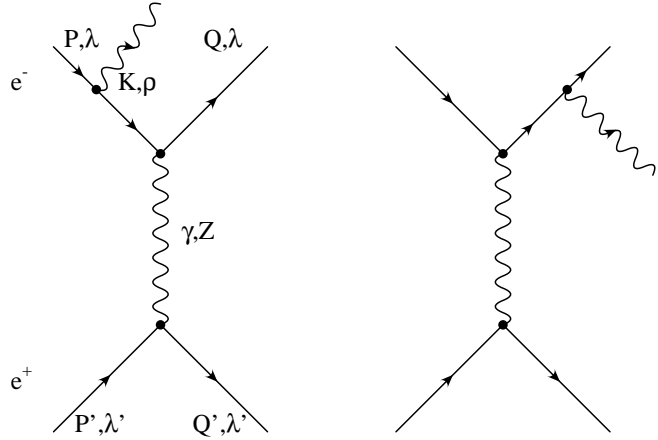,height=6cm}
 \caption{The t-channel tree level Feynman diagrams contributing to the ${\cal O}(\alpha)$ single bremsstrahlung cross section for initial and final state electron line emission.}
 \label{fig:Atrlvl}
\end{figure}
\end{center}

\noindent The ${\cal O}(\alpha)$ t-channel single bremsstrahlung electron-line emission amplitudes are shown in Fig. \ref{fig:Atrlvl}.
In the high-energy limit $\frac{{m_e}^2}{-t} \ll 1$, one can legitimately make use of the massless spinor formulation described in chapter \ref{sec:defid} and of the helicity conserving "magic" polarization-vector in Eq. \ref{eq:epsmu}.  
The tree-level amplitude, from now on denoted by $ A^{TL}\!\!\!\!\!\!_{e^{-}}$, can, for initial state electron-line emission, Fig. \ref{fig:Atrlvl}, be expressed as

\begin{eqnarray}
A^{TL}\!\!\!\!\!\!_{e^{-}} &=& \frac{i\,e^3}{-2PK} \langle Q',\lambda'|\, \gamma_{\mu} \, |P',\lambda' \rangle G_{\lambda,\lambda\!'}(t') \times \nonumber \\
& &  \langle Q,\lambda|\gamma^{\mu}(\rlap/ \!P - \rlap/ \!K) \, \rlap/ \varepsilon \, |P,\lambda  \rangle  \label{eq:Born1} \\[0.2cm]
&=& \frac{i\,e^3 G_{\lambda,\lambda\!'}(t') \delta_{\lambda,-\rho}}{\langle K,P \rangle_{\lambda} \langle K,P \rangle_{-\lambda}} 
\; [ \; 2P\varepsilon \langle Q',\lambda'|\, \gamma_{\mu} |P',\lambda' \rangle \langle Q,\lambda| \, \gamma^{\mu} |P,\lambda  \rangle \nonumber \\[0.2cm]
& & + \;\;  2 \, \langle Q',\lambda'|\, \rlap/ \varepsilon \, |P',\lambda' \rangle \langle Q,\lambda| \, \rlap/ \!K |P,\lambda  \rangle \; ] \\[0.2cm] \nonumber
&=& \frac{i\,e^3 \sqrt{8} \rho G_{\lambda,\lambda\!'}(t') \delta_{\lambda,-\rho}}{\langle K,P \rangle_{\lambda} \langle K,P \rangle_{-\lambda} \langle h,K \rangle_{\rho}} \times \\[0.2cm] \nonumber                                       & & \left[ \;\;\; \langle Q,P \rangle_{-\lambda} \langle P,K \rangle_{\lambda} \left\{ \begin{array}{ll} \!\!\! \langle Q',Q \rangle_{-\lambda'} \langle P,P' \rangle_{\lambda'} & \!\!\!\! \mbox{$, \lambda=\lambda'$} \\ \nonumber
\!\!\! \langle Q',P \rangle_{-\lambda'} \langle Q,P' \rangle_{\lambda'} & \!\!\!\! \mbox{$, \lambda=-\lambda'$} \end{array} \right. \right. \\[0.2cm] \nonumber
& & \left. + \,\, \langle Q,K \rangle_{-\lambda} \langle K,P \rangle_{\lambda} \left\{ \begin{array}{ll} \!\!\! \langle Q',Q \rangle_{-\lambda'} \langle K,P' \rangle_{\lambda'} & \!\!\!\! \mbox{$, \lambda=\lambda'$} \\ \nonumber
\!\!\! \langle Q',K \rangle_{-\lambda'} \langle Q,P' \rangle_{\lambda'} & \!\!\!\! \mbox{$, \lambda=-\lambda'$} \end{array} \right. \right] \\[0.2cm] \nonumber
&=& \frac{i\,e^3 \sqrt{8} \rho G_{\lambda,\lambda\!'}(t') \delta_{\lambda,-\rho}}{\langle P,K \rangle_{-\lambda} \langle h,K \rangle_{\rho}} \times \\[0.2cm] \nonumber                                                              & & \left[ \;\;\; \langle Q,P \rangle_{-\lambda} \left\{ \begin{array}{ll} \!\!\! \langle Q',Q \rangle_{-\lambda'} \langle P,P' \rangle_{\lambda'} & \!\!\!\! \mbox{$, \lambda=\lambda'$} \\ \nonumber
\!\!\! \langle Q',P \rangle_{-\lambda'} \langle Q,P' \rangle_{\lambda'} & \!\!\!\! \mbox{$, \lambda=-\lambda'$} \end{array} \right. \right. \\[0.2cm] \nonumber
& & \left. - \,\, \langle Q,K \rangle_{-\lambda} \left\{ \begin{array}{ll} \!\!\! \langle Q',Q \rangle_{-\lambda'} \langle K,P' \rangle_{\lambda'} & \!\!\!\! \mbox{$, \lambda=\lambda'$} \\ \nonumber
\!\!\! \langle Q',K \rangle_{-\lambda'} \langle Q,P' \rangle_{\lambda'} & \!\!\!\! \mbox{$, \lambda=-\lambda'$} \end{array} \right. \right] \\[0.2cm] \nonumber
&=& \frac{i\,e^3 \sqrt{8} \rho G_{\lambda,\lambda\!'}(t') \delta_{\lambda,-\rho}}{\langle P,K \rangle_{\rho} \langle h,K \rangle_{\rho}} \times \\[0.2cm] \nonumber                                                              & & \left\{ \begin{array}{ll} \!\!\!\! \langle Q',Q \rangle_{\rho} ( \langle Q,P \rangle_{-\lambda} \langle P,P' \rangle_{\lambda} - \langle Q,K \rangle_{-\lambda} \langle K,P' \rangle_{\lambda} ) & \!\!\!\! \mbox{$, \lambda=\lambda'$}  \\ \nonumber
\!\!\!\! \langle Q,P' \rangle_{\rho} ( \langle Q,P \rangle_{-\lambda} \langle Q',P \rangle_{\lambda} - \langle Q,K \rangle_{-\lambda} \langle Q',K \rangle_{\lambda} ) & \!\!\!\! \mbox{$, \lambda=-\lambda'$} \end{array} \right. \\[0.2cm] \nonumber
&=& \frac{i\,e^3 \sqrt{8} \rho G_{\lambda,\lambda\!'}(t') \delta_{\lambda,-\rho}}{\langle P,K \rangle_{\rho} \langle h,K \rangle_{\rho}} \! \left\{ \begin{array}{ll} \!\!\!\! \langle Q',Q \rangle_{\rho} \langle Q,Q' \rangle_{-\lambda} \langle Q',P' \rangle_{\lambda} & \!\!\!\! \mbox{$, \lambda=\lambda'$} \\ \nonumber 
\!\!\!\! \langle Q,P' \rangle_{\rho} \langle Q,P' \rangle_{-\lambda} \langle P',Q' \rangle_{\lambda} & \!\!\!\! \mbox{$, \lambda=-\lambda'$} \end{array} \right. \nonumber \\[0.2cm]
&=& i\,e^3 \sqrt{8} \, \lambda' \,G_{\lambda,\lambda\!'}(t')\, \frac{ \langle h',\widehat{h'} \rangle_{-\rho} ( \langle h,h'\rangle_{\rho})^2}{ \langle h,K \rangle_{\rho}  \langle \widehat{h},K \rangle_{\rho}} \;, \label{eq:Born2} 
\end{eqnarray}

\noindent where the properties of $\varepsilon_{\mu}$ of Eq. \ref{eq:epsmu}, four-momentum conservation and the Fierz-identity Eq. \ref{eq:Fierz} were used. Eq. \ref{eq:Born2}, derived for a specific electron-photon helicity correlation, holds in fact for both helicity cases. This was achieved by using the helicity dependent $h$-spinors of chapter \ref{sec:defid}. Although Eq. \ref{eq:Born2} represents the numerically most stable form for $A^{TL}\!\!\!\!\!\!_{e^{-}}$, there is another way of writing \ref{eq:Born1} by means of the identity \ref{eq:id3} which will prove to be very useful in the calculation at a later point and is indeed the main application of \ref{eq:id3}:

\begin{eqnarray}
A^{TL}\!\!\!\!\!\!_{e^{-}} &=& \frac{ie^3}{\widehat{h}K} \,G_{\lambda,\lambda\!'}(t')\, [\frac{\rho \lambda t - 2hK}{t} \langle Q',\lambda'|\, \gamma_{\mu}
\, |P',\lambda' \rangle \langle Q,\lambda|\gamma^{\mu} \, |P,\lambda  \rangle \nonumber \\
& & + \, \frac{2}{t} \, \langle Q',\lambda'|\, \rlap/ \!h \, |P',\lambda' \rangle \langle Q,\lambda| \, \rlap/ \!K \, |P,\lambda  \rangle ] \, \widehat{h} \varepsilon  \label{eq:Born3}
\end{eqnarray}

\noindent Eq. \ref{eq:Born3} will play an important role in separating out the
"true" tree-level part of the complete ${\cal O}(\alpha^2)$ internal emission 
calculation in chapter \ref{sec:intem}; see also Ref. \cite{thesis}.

\chapter{The ${\cal O}(\alpha^{2})$ Single Bremsstrahlung Amplitudes} \label{sec:singBramp}

\noindent This chapter is devoted to describing in detail the steps of the 
second order calculation. While a lot of analytical work is presented, 
all occuring loop integrals were evaluated in the appendices of Ref. \cite{thesis}
with the algebraic manipulation language
FORM.
It is assumed from this point on that all UV-divergences will be properly 
taken care of by the appropriate counter terms in the Lagrangian such that in 
the on-shell renormalization scheme $-e$ and $m_e$ correspond to the physically
observable charge and mass of the electron respectively.

\begin{center}
\begin{figure}
 \epsfig{file=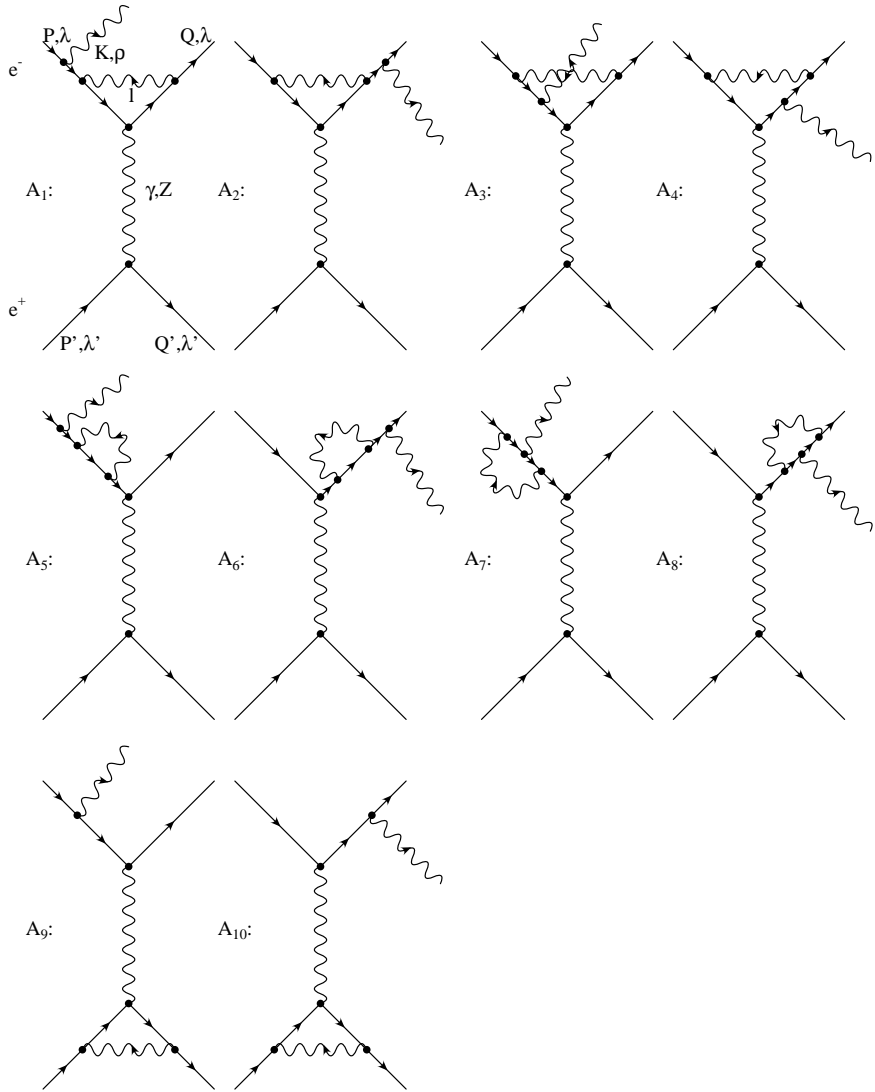,height=14.5cm}
 \caption{The one particle t-channel exchange Feynman 
diagrams, modulo vacuum polarization graphs, for initial and final state 
electron line emission that give a nonzero contribution to the ${\cal O}(\alpha^2)$ 
single bremsstrahlung cross section after renormalization is carried out in 
the chosen on-shell scheme. Also Z-exchange can be included.}
 \label{fig:fdvirt}
\end{figure}
\end{center}

\noindent Employing standard Feynman techniques \cite{ryder,pokorski} gives 18 electron 
line emission graphs
for the desired level of precision in the previously described context 
(compare with chapter \ref{sec:intro}). 
The single bremsstrahlung amplitudes contributing to ${\cal O}(\alpha^{2})$ radiative
corrections to the low angle Bhabha cross section are listed in 
Fig. \ref{fig:fdvirt} for 
t-channel electron-line emission. Positron-line emission can be obtained from 
these amplitudes through crossing relations:

\begin{eqnarray}
P \; & \longleftrightarrow & - \; Q' \label{eq:cross2a} \\
P' & \longleftrightarrow & - \; Q  \label{eq:cross2b} \\
\lambda \;\; & \longrightarrow & - \; \lambda \label{eq:cross2c}
\end{eqnarray}

\noindent A crossing rule can also be used to calculate the initial state 
s-channel contribution by applying \ref{eq:cross2b} to the electron-line 
t-channel calculation. The s-channel final state contribution is attained by
employing \ref{eq:cross2a} and \ref{eq:cross2c} to the expressions  
corresponding to the diagrams shown in Fig. \ref{fig:fdvirt}.
All vacuum polarization graphs that would in principle also belong to
this class of diagrams are explicitly omitted here, since they are 
already included in BHLUMI4.01 \cite{JadWas95}.

\newpage
\noindent Following the notation of Fig. \ref{fig:fdvirt} the various graphs
translate into the amplitudes given by the expressions listed below. It should
be noted that the Feynman gauge \cite{kaku,pokorski} was picked to arrive at the following
representation. The divergent integrals are regularized by means of the
n-dimensional regularization technique.

\vspace{-0.36cm}
\begin{eqnarray}
A_1 &=& \frac{e^5}{(2\pi)^4} \langle Q',\lambda'| \, \gamma_{\mu} \, |P',\lambda' \rangle G_{\lambda,\lambda\!'}(t') \times \nonumber \\
& & \int \frac{\mu^{\epsilon}d^nl}{-2PK} \frac{\langle Q,\lambda| \, \gamma_{\nu}(\not l +
  \rlap/ \!Q) \gamma^{\mu} (\not l + \rlap/ \!P - \rlap/ \!K) \gamma^{\nu}(\rlap/ \!P - \rlap/ \!K) \, \rlap/ \varepsilon \, |P,\lambda \rangle}
  {(l^2 - {m_0}^2)((l+P-K)^2 - {m_e}^2)((l+Q)^2 - {m_e}^2) + i\epsilon} \nonumber  \\
& & \label{eq:A1} \\
A_2 &=& \frac{e^5}{(2\pi)^4}  \langle Q',\lambda'| \, \gamma_{\mu} \, |P',\lambda' \rangle G_{
\lambda,\lambda\!'}(t') \times \nonumber \\
& & \int \frac{\mu^{\epsilon}d^nl}{2QK} \frac{ \langle Q,\lambda| \,\rlap/ \varepsilon \, (\rlap/ \!Q + \rlap/ \!K) \gamma_{\nu}(\not l + \rlap/ \!Q + \rlap/ \!K) \gamma^{\mu} (\not l + \rlap/ \!P) \gamma^{\nu} \, |P,\lambda \rangle}{(l^2 - {m_0}^2)((l+P)^2 - {m_e}^2)((l+Q+K)^2 - {m_e}^2) + i\epsilon} \nonumber  \\
& & \label{eq:A2} \\
A_3 &=& \frac{e^5}{(2\pi)^4}   \langle Q',\lambda'| \, \gamma_{\mu} \, |P',\lambda' \rangle G_{
\lambda,\lambda\!'}(t') \times \nonumber \\
&&\!\!\!\!\!\!\!\!\!\! \int \frac{d^4l \langle Q,\lambda| \, \gamma_{\nu}(\not l +
  \rlap/ \!Q) \gamma^{\mu} (\not l + \rlap/ \!P - \rlap/ \!K) \, \rlap/ 
\varepsilon \, (\not l + \rlap/ \!P) \gamma^{\nu} \, |P,\lambda \rangle}
{(l^2 - {m_0}^2)((l+P)^2 - {m_e}^2)((l+P-K)^2 - {m_e}^2)((l+Q)^2 - {m_e}^2) 
+ i\epsilon} \nonumber \\
& & \label{eq:A3} \\
A_4 &=& \frac{e^5}{(2\pi)^4}  \langle Q',\lambda'| \, \gamma_{\mu} \, |P',\lambda' \rangle G_{\lambda,\lambda\!'}(t') \times \nonumber \\
&&\!\!\!\!\!\!\!\!\!\! \int \frac{d^4l  \langle Q,\lambda| \, \gamma_{\nu}(\not l +\rlap/ \!Q) \, \rlap/ \varepsilon \, (\not l + \rlap/ \!Q + \rlap/ \!K)  \gamma^{\mu} (\not l + \rlap/ \!P) \gamma^{\nu} \, |P,\lambda \rangle}{(l^2 - {m_0}^2)((l+P)^2 - {m_e
}^2)((l+Q+K)^2 - {m_e}^2)((l+Q)^2 - {m_e}^2) + i\epsilon} \nonumber \\
& & \label{eq:A4} \\
A_5 &=& \frac{e^5}{(2\pi)^4}  \langle Q',\lambda'| \, \gamma_{\mu} |P',\lambda' \rangle G_{\lambda,\lambda\!'}(t') \times \nonumber \\
& &\!\!\!\!\!\!\!\!\!\!\!\!\!\!\!\!\!\!\!\!\!\!\!  \langle Q,\lambda| \, \gamma^{\mu}\frac{\rlap/ \!P - \rlap/ \!K}{-2PK} \int \mu^{\epsilon}d^nl
\frac{\gamma_{\nu}(\not l + \rlap/ \!P - \rlap/ \!K)\gamma^{\nu}}{(l^2-{m_0}^2)((l+P-K)^2-{m_e}^2) + i\epsilon} \, \frac{\rlap/ \!P - \rlap/ \!K}{-2PK}\, \rlap/ \varepsilon \, |P,\lambda \rangle  \nonumber \\
& & \label{eq:A5} \\
A_6 &=& \frac{e^5}{(2\pi)^4}  \langle Q',\lambda'| \, \gamma_{\mu} |P',\lambda' \rangle G_{\lambda,\lambda\!'}(t') \times \nonumber \\
& &\!\!\!\!\!\!\!\!\!\!\!\!\!\!\!\!\!\!\!\!\!\!\!  \langle Q,\lambda| \, \rlap/ \varepsilon \, \frac{\rlap/ \!Q + \rlap/ \!K}{2QK} 
\int \mu^{\epsilon}d^nl \frac{\gamma_{\nu}(\not l + \rlap/ \!Q + \rlap/ \!K)\gamma^{\nu}}
{(l^2-{m_0}^2)((l+Q+K)^2-{m_e}^2) + i\epsilon} \, \frac{\rlap/ \!Q + \rlap/ \!K}{2QK}\gamma^{\mu} |P,\lambda \rangle \nonumber \\
& & \label{eq:A6} \\
A_7 &=& \frac{e^5}{(2\pi)^4}  \langle Q',\lambda'| \, \gamma_{\mu} |P',\lambda' \rangle G_{\lambda,\lambda\!'}(t') \times \nonumber \\
& &\!\!\!\!\!\!\!\!\!\!\!\!\!\!\!\!\!\!\!\!\!\!\!  \langle Q,\lambda| \, \gamma^{\mu}\frac{\rlap/ \!P - \rlap/ \!K}{-2PK} \! \int 
\!\! \frac{\mu^{\epsilon}d^nl \; \gamma_{\nu}(\not l + \rlap/ \!P - \rlap/ 
\!K)\, \rlap/ \varepsilon \,(\not l + \rlap/ \!P)\gamma^{\nu}}
{(l^2-{m_0}^2)((l+P)^2-{m_e}^2)((l+P-K)^2-{m_e}^2) + i\epsilon} |P,\lambda 
\rangle \nonumber \\
& & \label{eq:A7} \\
A_8 &=& \frac{e^5}{(2\pi)^4} \langle Q',\lambda'| \, \gamma_{\mu} |P',\lambda' \rangle G_{\lambda,\lambda\!'}(t') \times \nonumber \\
& &\!\!\!\!\!\!\!\!\!\!\!\!\!\!\!\!\!\!\!\!\!\!\!  \langle Q,\lambda| \! \int \!\! \frac{\mu^{\epsilon}d^nl \; \gamma_{\nu}(\not l + \rlap/ \!Q)\, \rlap/ \varepsilon \,(\not l + \rlap/ \!Q + \rlap/ \!K)\gamma^{\nu}}{(l^2-{m_0}^2)((l+Q)^2-{m_e}^2)((l+Q+K)^2-{m_e}^2) + i\epsilon} \, \frac{\rlap/ \!Q + \rlap/ \!K}{2QK} \gamma^{\mu} |P,\lambda \rangle \nonumber \\[0.2cm]
& & \label{eq:A8} \\
A_9 &=& \frac{e^5}{(2\pi)^4}  \langle Q,\lambda| \, \gamma_{\mu} \frac{\rlap/ \!P - \rlap/ \!K}{-2PK} \, \rlap/ \varepsilon \, |P,\lambda \rangle G_{\lambda,\lambda\!'}(t') \times \nonumber \\ 
& & \int \frac{\mu^{\epsilon}d^nl  \langle Q',\lambda'| \, \gamma_{\nu}(\not l + \rlap/ \!Q') \gamma^{\mu} (\not l + \rlap/ \!P') \gamma^{\nu} |P',\lambda' \rangle}{(l^2 - {m_0}^2)((l+P')^2 - {m_e}^2)((l+Q')^2 - {m_e}^2) + i\epsilon} \nonumber \\
& & \label{eq:A9} \\
A_{10} &=& \frac{e^5}{(2\pi)^4}  \langle Q,\lambda| \, \rlap/ \varepsilon \, \frac{\rlap/ \!Q + \rlap/ \!K}{2QK} \gamma_{\mu} \, |P,\lambda \rangle G_{\lambda,\lambda\!'}(t') \times 
\nonumber \\
& & \int \frac{\mu^{\epsilon}d^nl \langle Q',\lambda'| \, \gamma_{\nu}(\not l + \rlap/ \!Q') \gamma^{
\mu} (\not l + \rlap/ \!P') \gamma^{\nu} |P',\lambda' \rangle}{(l^2 - {m_0}^2)((l+P')^
2 - {m_e}^2)((l+Q')^2 - {m_e}^2) + i\epsilon} \nonumber \\
& & \label{eq:A10} 
\end{eqnarray}

\noindent The following sections of this chapter will now go about solving 
the integrals appearing in Eqs. \ref{eq:A1} $-$ \ref{eq:A10} and give concise 
expressions for the corresponding amplitudes.

\section{Vertex Corrections} \label{sec:vertfun}

\noindent The name "vertex corrections" to the ${\cal O}(\alpha)$ single 
bremsstrahlung cross section implies that some kind of a 
factorization takes place, that separates the lower order amplitudes from 
the additional virtual corrections. This, however, is not true in general
and only holds in this calculation for $A_1$, $A_2$, $A_9$ and $A_{10}$
because of the application of the 
high energy approximation and the use of the "magic" polarization vector 
\ref{eq:eps} and the "magic" choices given in Eqs. \ref{eq:hdef} and  
\ref{eq:h'def}. 
The graphs calculated in this section include amplitudes \ref{eq:A1}, 
\ref{eq:A2}, \ref{eq:A7}, \ref{eq:A8}, \ref{eq:A9} and \ref{eq:A10}.
For the amplitudes $A_7$ and $A_8$, the described factorization does not 
take place despite the tools mentioned above. It still will be useful
in this case to use the broad headline "vertex correction" because of two
reasons:

\noindent The result for $A_7$ and $A_8$ is completely determined by form
factors derived in this section and secondly, the terms not proportional
to the lower order, tree level, amplitude turn out to be canceled exactly
by a contribution contained in the rather complicated expression for
$A_3$ and $A_4$, to be discussed in section \ref{sec:intem}. This 
circumstance will insure that all leading and subleading contributions
will be concentrated in the factorized part of the complete answer.

\noindent All results employ the on-shell renormalization scheme, i.e. 
they are
renormalized such that the vertex correction contribution vanishes if
both fermion legs are on the mass-shell and for zero four-momentum transfer.

\subsection{Scalar Integrals} \label{sec:vertscalint}

\noindent The Scalar integrals, in terms of which the following results of 
the various amplitudes are given, are defined in this section as follows:

\begin{eqnarray}
{B_{12}\!\!\!\!\!^{\alpha}} \,  &=&  B_0({m_e}^2 \! -\alpha;m_0,m_e) 
\label{eq:B12al} \\
B_{23} &=&  B_0(2{m_e}\!^2 \! +t';m_e,m_e)    \label{eq:B23} \\
{B_{23}\!\!\!\!\!^{0}} \,\, &=&  B_0({m_0}\!^2;m_e,m_e)    \label{eq:B230} \\
{C_{123}\!\!\!\!\!\!\!^{\alpha}} \,\,\, &=&  C_0({m_e}\!^2 \! -\alpha,2{m_e}\!^2
\! +t',{m_e}\!^2;m_0,m_e,m_e) \label{eq:C123alz} \\
C_{1230} &=& C_0(m_e\!^2,m_0\!^2,m_e\!^2,m_0,m_e,m_e) \label{eq:C1230}
\end{eqnarray}

\noindent It must be emphasized that the notation for scalar integrals in 
\ref{eq:B12al} $-$ \ref{eq:C1230} only refers to loop integrals with no
more than three internal propagators. This was done in order to have 
agreement with the notation implicitly employed by FORM in the appendices
of Ref. \cite{thesis}.

\noindent Eq. \ref{eq:C1230} is somewhat special among the integrals denoted
above in that it can be expressed analytically and without renormalization and
scale dependence:

\begin{eqnarray}
C_{1230} &=& \frac{1}{i\pi^2} \int d^4l \frac{1}{(l^2-m_0\!^2)(l^2+2lP)^2 
+ i \epsilon } \nonumber \\
&=& \frac{2}{i\pi^2} \int d^4l \int_0^1 dx \frac{x}{(l^2+2lPx -m_0\!^2(1-x)+ i \epsilon)^3} \nonumber \\
&=& - \int_0^1 dx \frac{x}{m_e\!^2x^2 - m_0\!^2 x + m_0\!^2 + i \epsilon} \nonumber \\ 
&=& - \frac{1}{2m_e\!^2} ln (m_e\!^2x^2 - m_0\!^2 x + m_0\!^2 + i \epsilon) \Big\vert_0^1
\nonumber \\
&=& m_e\!^{-2} \; ln \; \frac{m_0}{m_e} \; , \label{eq:C1230res} 
\end{eqnarray}

\noindent where standard Feynman parameters \cite{pokorski,ryder}, integral tables \cite{grad}
and formulas for integrals over four-dimensional Minkowski space \cite{pokorski,ryder,ramond} 
were used. In addition it follows from \ref{eq:C1230} that $P^2=m_e\!^2$, so
that only masses are parameters of this integral.  
The significance of \ref{eq:C1230res} here is merely a notational
one. It means that $m_e\!^2 C_{1230}$-terms are actually of order 
${\cal O}(m_e\!^0)$ and need to be taken into account; it also is IR-divergent,
which implies the possibility of a strong analytical check of the expected
limit for $m_0 \rightarrow 0$; see chapter \ref{sec:YFS} for an analysis of
this problem. The same behaviour with respect to a possible $m_e\!^{-2}$ 
dependence is not given for the other integrals listed. Dimensional reasons
exclude it for $A_0$ and $B_0$ functions and numerical checks for the
other integrals in the regime where products of four-momenta are much
larger than $m_e\!^{2}$.

\subsection{The Vertex Function}

\noindent The vertex function ${\Gamma^{\mu}\!\!\!\!_o}\,((P-Q)^2,P^2,Q^2)$ is given by the integral:

\begin{equation}
{\Gamma^{\mu}\!\!\!\!_o} \, = \, \frac{1}{i\pi^2} \int \mu^{\epsilon}d^nl \frac{\gamma_{\nu}(\not l +
  \rlap/ \!Q + m_e) \gamma^{\mu} (\not l + \rlap/ \!P + m_e) \gamma^{\nu}}
  {(l^2 - {m_0}^2)((l+P)^2 - {m_e}^2)((l+Q)^2 - {m_e}^2) + i\epsilon} \label{eq:vertexf}
\end{equation}

\noindent The numerator of the integral \ref{eq:vertexf} can be simplified by means of standard
Dirac-algebra manipulations as follows:

\begin{eqnarray}
&&
\gamma_{\nu}(\not l + \rlap/ \!Q + m_e) \gamma^{\mu} (\not l + \rlap/ \!P + m_e) \gamma^{\nu} 
\nonumber \\
&=& \gamma_{\nu}(\not l + \rlap/ \!Q)\gamma^{\mu}(\not l + \rlap/ \!P)\gamma^{\nu} 
+m_e\gamma_{\nu}\gamma^{\mu}(\not l + \rlap/ \!P) \gamma^{\nu}
+m_e\gamma_{\nu}(\not l + \rlap/ \!Q) \gamma^{\mu}\gamma^{\nu}
 \nonumber \\ && + {m_e}^2\gamma_{\nu}\gamma^{\mu}\gamma^{\nu} \nonumber \\
&=& -2(\not l + \rlap/ \!P) \gamma^{\mu} (\not l + \rlap/ \!Q)
+4m_e(l + P)^{\mu} + 4m_e(l + Q)^{\mu} - 2{m_e}^2\gamma^{\nu} \nonumber \\
&=& -2((-\gamma^{\mu}(\not l + \rlap/ \!P) + 2(l + P)^{\mu})(\not l + \rlap/ \!Q)
+  (m_e-terms) \nonumber \\
&=& -4(\not l + \rlap/ \!Q)(l + P)^{\mu} + 2\gamma^{\mu}(-(\not l + \rlap/ \!Q)(\not l + \rlap/ \!P) +2(l + P)(l + Q)) \nonumber \\ && + (m_e-terms) \nonumber \\
&=& -4(\not l + \rlap/ \!Q)(l + P)^{\mu} + 4\gamma^{\mu}(l + P)(l + Q)
+2((\not l + \rlap/ \!Q)\gamma^{\mu} \nonumber \\ && - 2(l + Q)^{\mu})(\not l + \rlap/ \!P)
+ (m_e-terms) \nonumber \\
&=& 2(\not l + \rlap/ \!Q)\gamma^{\mu}(\not l + \rlap/ \!P) + 4\gamma^{\mu}
(l + P)(l + Q) - 4(l + P)^{\mu}(\not l + \rlap/ \!Q) \nonumber \\ && - 4(l + Q)^{\mu}(\not l + \rlap/ \!P) 
+ (m_e-terms) \nonumber \\
&=& 2(\not l + \rlap/ \!Q - m_e)\gamma^{\mu} (\not l + \rlap/ \!P - m_e)
+2m_e\gamma^{\mu} (\not l + \rlap/ \!P) + 2m_e(\not l + \rlap/ \!Q) \gamma^{\mu}
 \nonumber \\ && + 4\gamma^{\mu}(l + P)(l + Q)
 - 4(l + P)^{\mu}(\not l + \rlap/ \!Q)
- 4(l + Q)^{\mu}(\not l + \rlap/ \!P)  \nonumber \\ && + 4m_e(l + P)^{\mu} + 4m_e(l + Q)^{\mu}
-4{m_e}^2\gamma^{\nu} \nonumber \\
&=& 2(\not l + \rlap/ \!Q - m_e)\gamma^{\mu}  (\not l + \rlap/ \!P - m_e) 
+ 2m_e\gamma^{\mu}  (\not l + \rlap/ \!P - m_e)  \nonumber \\ && + 2m_e(\not l + \rlap/ \!Q - m_e) \gamma^{\mu} 
+ 4\gamma^{\mu}(l + P)(l + Q) - 4(l + P)^{\mu}(\not l + \rlap/ \!Q - m_e) \nonumber \\ &&
- 4(l + Q)^{\mu}(\not l + \rlap/ \!P - m_e) \nonumber \\
&=& 4\gamma^{\mu}(l + P)(l + Q) + 4m_el^{\mu} - 4\not l  (2l + P + Q)^{\mu}
 -2\gamma^{\mu}l^2  \nonumber \\ && + 4l^{\mu}(\not l + \rlap/ \!Q - m_e)
-2(\rlap/ \!Q - m_e)\not l\gamma^{\mu} + 2\not l\gamma^{\mu}(\rlap/ \!P - m_e)
\nonumber \\ && + 2(\rlap/ \!Q - m_e)\gamma^{\mu}(\rlap/ \!P - m_e)  + 2m_e\gamma^{\mu}(\rlap/ \!P - m_e)
+ 2m_e(\rlap/ \!Q - m_e)\gamma^{\mu} \nonumber \\ && - 4(l + P)^{\mu}(\rlap/ \!Q - m_e)
-4(l + Q)^{\mu}(\rlap/ \!P - m_e) \nonumber \\
&=& 4\gamma^{\mu}(PQ + l(P + Q) + \frac{l^2}{2}) - 4\not l(P + Q + l)^{\mu} + 4m_el^{\mu} \nonumber \\
& & + (\rlap/ \!Q - m_e)(2m_e\gamma^{\mu} - 2\not l\gamma^{\mu} - 4(l + P)^{\mu}
+ 4l^{\mu}) \nonumber \\
& & +(2m_e\gamma^{\mu} + 2\not l\gamma^{\mu} - 4(l + Q)^{\mu})(\rlap/ \!P - m_e)
\nonumber \\
& & + 2(\rlap/ \!Q - m_e)\gamma^{\mu}(\rlap/ \!P - m_e) \nonumber \\
&=& 4\gamma^{\mu}(PQ + l(P + Q) + \frac{l^2}{2}) - 4\not l(P + Q +l
)^{\mu} + 4m_el^{\mu} \nonumber \\
& & + (\rlap/ \!Q - m_e)(2m_e\gamma^{\mu} + 2\gamma^{\mu}\not l - 4(l + P)^{\mu})
\nonumber \\
& & +(2m_e\gamma^{\mu} + 2\not l\gamma^{\mu} - 4(l + Q)^{\mu})(\rlap/ \!P - m_e)
\nonumber \\
& & + 2(\rlap/ \!Q - m_e)\gamma^{\mu}(\rlap/ \!P - m_e)  \label{eq:numvert}
\end{eqnarray}

\noindent The first line of Eq. \ref{eq:numvert} contributes regardless of whether the fermion lines are on- or off-shell and can be found in standard textbooks, for example in Ref. \cite{itzzub}.
In case one fermion line is on- the other one off-shell, either the second or third line, respectively, contributes in addition.
The last term in Eq. \ref{eq:numvert} is only needed for the case where both $P$ and
$Q$ are off-shell and thus will not enter into the present ${\cal O}(\alpha^2)$ level
calculation. The mass-terms are retained only until the last stage of the FORM-reduction, after which only terms of ${\cal O}({m_e}^0)$ are kept; see the $f\!fac$-routines in Ref. \cite{thesis} for details.

\noindent For initial state electron-line radiation, Eqs. \ref{eq:A1}, \ref{eq:A9}, the renormalized vertex function $\Gamma^{\mu}$ can be written as 

\begin{equation}
\Gamma^{\mu} \; = \; F\!F \gamma^{\mu} + F\!F_a (\rlap/ \!P - \rlap/ \!K)(P - K)^{\mu} + F\!F_b (\rlap/ \!P - \rlap/ \!K) Q^{\mu} \label{eq:Gdecomp}
\end{equation}

\noindent With the previously defined notation for scalar integrals in section \ref{sec:vertscalint} for the external emission graphs $A_1$ and $A_9$, the decomposition \ref{eq:Gdecomp} is done by the FORM-routine $f\!fac$, Ref. \cite{thesis}, and yields the following
results for the various form-factor terms up to ${\cal O}({m_e}^0)$:
 
\begin{eqnarray}
F\!F &=& -2 - 4 m_e\!^2 C_{1230} - 2 t' C_{123}\!\!\!\!\!\!\!^{\alpha} \,\,\, + {B^0}\!\!\!_{23} + 2 B_{12}\!\!\!\!\!^{\alpha} - 3 B_{23} \nonumber \\
& & - \frac{3\alpha}{t'+\alpha}(B_{12}\!\!\!\!\!^{\alpha} - B_{23}) \label{eq:ffac} \\ 
F\!F_a &=& \frac{2}{t'+\alpha}(B_{12}\!\!\!\!\!^{\alpha} - B_{23}) \label{eq:ffa} \\
F\!F_b &=& - 4 {C_{123}\!\!\!\!\!\!\!^{\alpha}}  \,\,\, + \frac{4}{t'+\alpha} ({B^0}\!\!\!_{23} +\frac{3}{2} + \alpha {C_{123}\!\!\!\!\!\!\!^{\alpha}} \,\,\,\, ) \nonumber \\
& & + \frac{4}{(t'+\alpha)^2}(t' B_{12}\!\!\!\!\!^{\alpha} - \frac{3}{2} \alpha B_{12}\!\!\!\!\!^{\alpha} \, + \frac{1}{2} \alpha B_{23} - 2 t' B_{23}) \;, \label{eq:ffb}
\end{eqnarray}

\noindent for the off-shell graph $A_1$ and

\begin{eqnarray}
\!\!\!\!\!\!\!\!\!\!\!\!\!\!\!\!\!\!\!\!\!\!\!\!\!\!\!\!\!\!\!\!\!\!\!\!\!\!\!\!\!\!\!\!\!\!\!\!\!\!\!\!F\!F^0 &=& 2 -4 m_e\!^2 C_{1230} - 2 t' {C^0}\!\!\!_{123} +3{B^0}\!\!\!_{23}- 3 B_{23} \\
\!\!\!\!\!\!\!\!\!\!\!\!\!\!\!\!\!\!\!\!\!\!\!\!\!\!\!\!\!\!\!\!\!\!\!\!\!\!\!\!\!\!\!\!\!\!\!\!\!\!\!\!{F\!F^0}\!\!\!_{a} &=& 0 \\
\!\!\!\!\!\!\!\!\!\!\!\!\!\!\!\!\!\!\!\!\!\!\!\!\!\!\!\!\!\!\!\!\!\!\!\!\!\!\!\!\!\!\!\!\!\!\!\!\!\!\!\!{F\!F^0}\!\!\!_{b} &=& 0  \;,
\end{eqnarray}

\noindent for the on-shell graph $A_9$.
 
\noindent The identity $(\rlap/ \!P - \rlap/ \!K)\gamma^{\mu}(\rlap/ \!P - \rlap/ \!K) = \alpha \gamma^{\mu} + 2 (\rlap/ \!P - \rlap/ \!K) (P-K)^{\mu}$ was used in the above decomposition as well as the renormalization condition that $\Gamma^{\mu}$ vanish on-shell and for zero transfer, $t'=0$.
Due to the choice of the polarization-vector $\varepsilon^{\mu}$ the expression
$\langle Q,\lambda|\,\rlap/ \varepsilon \,|P,\lambda \rangle = 0$, which means that the form-factors $F\!F_a$ and $F\!F_b$ don't contribute to the aspired 
level of accuracy for $A_1(A_2)$ and $A_9(A_{10})$. They will, however, prove to be very useful for testing the internal consistency of the calculation, as will be discussed in chapter \ref{sec:intgaugevar} and do contribute to the amplitudes  \ref{eq:A7}, \ref{eq:A8}. While it is not directly apparent, it can easily be checked that all
renormalized form factors are indeed finite and scale independent. 
Ref. \cite{thesis} also demonstrates that the renormalization condition is met  
correctly.

\noindent Inserting Eq. \ref{eq:vertexf} into the expression for $A_1$, Eq. \ref{eq:A1}, and using the decomposition Eq. \ref{eq:Gdecomp} as well as the "magic" properties of $\varepsilon_{\mu}$ yields:

\begin{eqnarray}
A_1 &=& \frac{ie^5}{16\pi^2} \, \langle Q',\lambda'| \, \gamma_{\mu} \, |P',\lambda' \rangle G_{
\lambda,\lambda\!'}(t') \times \nonumber \\                                     & & \langle Q,\lambda| \, \Gamma^{\mu}((P'-Q')^2,(P-K)^2,{m_e}^2) \,
\frac{\rlap/ \!P - \rlap/ \!K}{-2PK} \, \rlap/ \varepsilon \, |P,\lambda \rangle \nonumber \\
&=& \frac{ie^5}{16\pi^2} \, \langle Q',\lambda'|\, \gamma_{\mu} \, |P',\lambda' \rangle G_{
\lambda,\lambda\!'}(t') \times \nonumber \\     
& & \{ \langle Q,\lambda|\gamma^{\mu}\,\frac{\rlap/ \!P - \rlap/ \!K}{-2PK} \, \rlap/ \varepsilon \, |P,\lambda \rangle \, F\!F \nonumber \\
& & + \langle Q,\lambda|(\rlap/ \!P - \rlap/ \!K)^2 \, \rlap/ \varepsilon \, |P,\lambda \rangle \,
(F\!F_a (P-K)^{\mu} + F\!F_b Q^{\mu}) \, \} \nonumber \\
&=& \frac{ie^5}{16\pi^2} \, \langle Q',\lambda'|\, \gamma_{\mu} \, |P',\lambda' \rangle G_{\lambda,\lambda\!'}(t') \times \nonumber \\
& & \langle Q,\lambda|\gamma^{\mu}\,\frac{\rlap/ \!P - \rlap/ \!K}{-2PK}\, \rlap/ \varepsilon \, |P,\lambda \rangle \, F\!F \nonumber \\
&=& \frac{e^2}{16\pi^2} \, A^{TL}\!\!\!\!\!\!_{e^{-}} \, F\!F((P'-Q')^2,(P-K)^2)  \,  \delta_{\rho,-\lambda} \nonumber \\
\end{eqnarray}
 
\noindent For $A_2$ one gets analogously:

\begin{equation}
\!\!\!\!\!\!\!\!\!\!\!\!\!\!\!\!\!\!\!\!\!\!\!\!\!\!\!\!\!\!A_2 \,\,\,\, = \,\,\,\, \frac{e^2}{16\pi^2} \, A^{TL}\!\!\!\!\!\!_{e^{-}} \, F\!F((P'-Q')^2,(Q+K)^2) \,  \delta_{\rho,\lambda} \\
\end{equation}

\noindent Since the tree-level amplitude $A^{TL}\!\!\!\!\!\!_{e^{-}}$ already contains the helicity dependence implicitly it is convenient to write:

\begin{equation}
\!\!\!\!\!\!\!\!\!\!\!\!\!\!\!\!\!\!\!\!\!\!\!\!\!\!\!\!\!\!\!\!\!\!\!\!\!\!A_1 + A_2 \,\,\,=\,\,\, \frac{e^2}{16\pi^2} \, A^{TL}\!\!\!\!\!\!_{e^{-}} \,F\!F
(t',\gamma) \;, \label{eq:a1a2res} \\
\end{equation}

\noindent with 

\begin{equation}
\gamma \,\,\,=\,\,\, \left\{ \begin{array}{ll} {m_e}^2 + 2QK & \mbox{$, \rho=\lambda$} \\
{m_e}^2 - 2PK & \mbox{$, \rho=-\lambda$} \end{array} \right. \\ \label{eq:gammaparam}
\end{equation}

\noindent For the on-shell form-factor amplitudes $A_9$ and $A_{10}$ the derivation is even more straightforward and gives:

\begin{equation}
\!\!\!\!\!\!\!\!\!\!\!\!\!\!\!\!\!\!\!\!\!\!\!\!\!\!\!\!\!\!\!\!\!\!\!\!\!\!\!\!A
_9 + A_{10} \,\,\,\, =\,\,\,\, \frac{e^2}{16\pi^2} \, A^{TL}\!\!\!\!\!\!_{e^{-}} \, F\!F^0(t') \label{eq:a9a10res} \\
\end{equation}

\noindent As can be seen from the definition of the scalar integrals for the various form-factors the relation between $F\!F$ and $F\!F^0$ is given by:

\begin{equation}
\!\!\!\!\!\!\!\!\!\!\!\!\!\!\!\!\!\!\!\!\!\!\!\!\!\!\!\!\!\!\!\!\!\!\!\!\!\!\!\!\!F\!F^0(t') \,\,\,=\,\,\, F\!F(t',\gamma={m_e}^2) \\
\end{equation}

\begin{center}
\begin{figure}
\epsfig{file=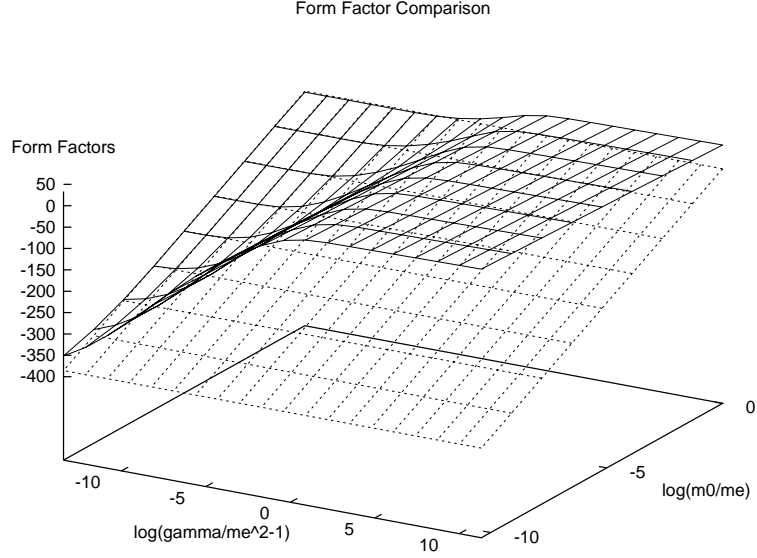,width=12cm}
\caption{The real parts of the on-shell (dashed lines) and off-shell form 
factors for various values of $\frac{\gamma}{m_e\!^2}-1$ at $t'=-1$. The off shell form factor is seen to approach $F\!F^0$ for $\gamma - m_e\!^2 \longrightarrow 0$, Eq. \protect\ref{eq:gammaparam}, and to have no $log \frac{m_0}{m_e}$ divergence for $\gamma \gg m_e\!^2$.}
\label{fig:ffac}
\end{figure}
\end{center}

\noindent This correspondence is shown to hold numerically in Fig. \ref{fig:ffac}.

\subsection{The Self-Energy Emission Amplitudes} \label{sec:selfenem}

\noindent The contribution from the off-shell self-energy emission amplitudes \ref{eq:A7}, \ref{eq:A8} can be described by the expressions for the off-shell form factors \ref{eq:ffac}, \ref{eq:ffa} and \ref{eq:ffb} in the limit of zero transfer $(t'=0)$ as follows:

\begin{eqnarray}
\varepsilon_{\mu} \, \Gamma^{\mu}(K^2,P^2,(P-K)^2)  
&=& F\!F  \, \rlap/ \varepsilon \, + \, \varepsilon_{\mu} P^{\mu} \,(F\!F_a+F\!F_b)\, (\rlap/ \!P - \rlap/ \!K) \nonumber \\ 
&=& (- 2 - 4m_e\!^2 C_{1230} +{B^0}\!\!\!_{23} - B_{12}\!\!\!\!\!^{\alpha} \,\,\, ) \, \rlap/ \varepsilon \, + \nonumber \\
& & \frac{\varepsilon_{\mu} P^{\mu}}{\alpha} \, (6 + 4 {B^0}\!\!\!_{23} - 4 B_{12}\!\!\!\!\!^{\alpha} \,\,\, ) \, (\rlap/ \!P - \rlap/ \!K) 
\end{eqnarray}

\noindent Using the above limit and decomposition gives for amplitude \ref{eq:A7}:

\begin{eqnarray} 
A_7 &=& \frac{e^2}{16\pi^2} \, (- 2 - 4m_e\!^2 C_{1230} + {B^0}\!\!\!_{23} -  B_{12}\!\!\!\!\!^{\alpha} \,\, ) \, A^{TL}\!\!\!\!\!\!_{e^{-}} \, +  \nonumber \\
&& \frac{ie^5}{16\pi^2} (6 + 4 {B^0}\!\!\!_{23} - 4 B_{12}\!\!\!\!\!^{\alpha} \,\,) 
\frac{\varepsilon_{\mu} P^{\mu}}{\alpha} \, G_{\lambda,\lambda\!'}(t')  \langle Q',\lambda'| \, \gamma_{\mu} \, |P',\lambda' \rangle \, \langle Q,\lambda| \, \gamma^{\mu} \, |P,\lambda \rangle \nonumber \\
&=& \frac{e^2}{16\pi^2} \, (- 2 - 4m_e\!^2 C_{1230} + {B^0}\!\!\!_{23} - B_{12}\!\!\!\!\!^{\alpha} \,\, ) \, A^{TL}\!\!\!\!\!\!_{e^{-}} \, + \nonumber \\
&& \frac{ie^5}{16\pi^2 \alpha} \, (6 + 4 {B^0}\!\!\!_{23} - 4 B_{12}\!\!\!\!\!^{\alpha} \,\,)
\left[\frac{2 \varepsilon_{\mu} P^{\mu} \delta_{\lambda,-\rho}}{t+\beta} G_{\lambda,\lambda\!'}(t') \right. \, \times \nonumber \\
&& \langle Q',\lambda'| \, \rlap/ \!Q \, |P',\lambda' \rangle \, \langle Q,\lambda| \, \rlap/ \!K \, |P,\lambda \rangle  \;\;
 - \left. \frac{\alpha t A^{TL}\!\!\!\!\!\!_{e^{-}}}{2ie^3(t+\beta)} \, \right] \nonumber \\
&=& \frac{e^2}{16 \pi^2} \, \left( \, \frac{t-\beta}{t+\beta} \, \left[B_{12}\!\!\!\!\!^{\alpha} \,\,\, -  {B^0}\!\!\!_{23} \; \right]\,\, - 4m_e\!^2 C_{1230} - \frac{5t+2\beta}{t+\beta} \, \right) \, A^{TL}\!\!\!\!\!\!_{e^{-}} \, + \nonumber \\ 
&& \frac{ie^5}{16 \pi^2} \, \frac{2 \varepsilon_{\mu} P^{\mu}}{(t+\beta)} 
\frac{\delta_{\lambda,-\rho}}{\alpha} \,
 (6 + 4 {B^0}\!\!\!_{23} -  4 B_{12}\!\!\!\!\!^{\alpha} \,\,\,) G_{\lambda,\lambda\!'}(t') \, \times \nonumber \\ 
&& \langle Q',\lambda'| \, \rlap/ \!Q \, |P',\lambda' \rangle \, \langle Q,\lambda| \, \rlap/ \!K \, |P,\lambda \rangle  \label{eq:A7res}
\end{eqnarray}

\noindent The important identity \ref{eq:Born3} was used in deriving the result in \ref{eq:A7res}, in which all terms proportional to $A^{TL}\!\!\!\!\!\!_{e^{-}}$ were identified. As will be discussed in more detail in chapter \ref{sec:llog}, terms multiplying the ${\cal O}(\alpha)$ single bremsstrahlung tree level amplitude contain the biggest corrections to the ${\cal O}(\alpha^2)$ cross section. The result for $A_8$ can be obtained from Eq. \ref{eq:A7res} through the crossing rule \ref{eq:cross1}:

\begin{equation}
A_8 \,\, = \,\, A_7(P \leftrightarrow -Q, \lambda \rightarrow -\lambda)
\label{eq:A8res}
\end{equation}

\section{The Electron Self-Energy Function} \label{sec:selfen}

\noindent The contribution of the amplitudes given by Eqs. \ref{eq:A5}, \ref{eq:A6} are determined by the solution of the electron self-energy function
 
\begin{eqnarray}
\Sigma_o(P,m_0,m_e) &=& \frac{1}{i\pi^2} \int \mu^{\epsilon}d^nl \frac{\gamma_{\nu}(\rlap/ \!P + \not l + m_e) \gamma^{\nu}}{(l^2-{m_0}^2)((P+l)^2-{m_e}^2)+i\epsilon} \label{eq:sigma0} \\[0.3cm]
&\equiv& {\Sigma_o}\!\!\!^a(P^2,m_0,m_e)\,m_e \, + \, {\Sigma_o}\!\!\!^b(P^2,m_0,m_e)\,\rlap/ \!P \label{eq:sigmadef}
\end{eqnarray}

\noindent The proper treatment of the mass terms in the expression \ref{eq:sigma0} for the self-energy contribution is essential here, since there are two independent renormalization constants in $\Sigma_o(P,m_0,m_e)$. In keeping with the hitherto adopted on-shell renormalization scheme, these two constants are fixed by imposing the renormalization conditions

\begin{eqnarray}
\Sigma(P,m_0,m_e)\big\vert_{\rlap/ \!P = m_e} &=& 0 \label{eq:renorm1} \\[0.2cm] 
\frac{\partial \Sigma(P,m_0,m_e)}{\partial P_{\mu}}\bigg\vert_{\rlap/ \!P = m_e} &=& 0 \label{eq:renorm2} 
\end{eqnarray}

\noindent The renormalization condition \ref{eq:renorm2} is the reason why the terms proportional to $m_e$ matter here: $\frac{\partial \Sigma^a}{\partial P_{\mu}}\big\vert_{\rlap/ \!P = m_e}$ contains terms proportional to $\frac{1}{m_e}$. 
After properly taking these into account, the FORM-routine selfen, Ref. \cite{thesis}, gives the following result: 

\begin{eqnarray}
\Sigma(P,m_0,m_e) &=&\left[4\left(B_{12}\!\!\!\!\!^{\alpha} \, - {B^0}\!\!\!_{12}\right) -4m_e\!^2 C_{1230} - 2\right]\,m_e \,\,+\nonumber \\       
&&  \left[4m_e\!^2 C_{1230} - \left(B_{12}\!\!\!\!\!^{\alpha}\;-{B^0}\!\!\!_{12}\,\right) \left(1 + \frac{{m_e}^2}{P^2} \right) -\frac{{m_e}^2}{P^2} + 3 \right]\, \rlap/ \!P \nonumber \\ 
&\approx& \left[4m_e\!^2 C_{1230} - \left(B_{12}\!\!\!\!\!^{\alpha}\; - {B^0}\!\!\!_{12}\;\right) +3\right]\, \rlap/ \!P \label{eq:sigma}
\end{eqnarray}

\noindent The mass-terms can be dropped in the final step in Eq. \ref{eq:sigma} since it represents the renormalized result. Doing the same at an earlier stage, however, would give the wrong $\Sigma^b(P^2,m_0,m_e)$-function.
Eq. \ref{eq:sigma} is also clearly finite and scale independent. The 
"$C_0$" function enters the self-energy result, a term with only two 
propagators, through the derivative in \ref{eq:renorm2} and is given 
explicitly in Eq. \ref{eq:C1230res}.
Inserting this result into Eq. \ref{eq:A5} yields:

\begin{eqnarray}
A_5 &=& \frac{ie^5}{16\pi^2}  \langle Q',\lambda'| \, \gamma_{\mu} |P',\lambda' \rangle G_{\lambda,\lambda\!'}(t') \times \nonumber \\
& &  \langle Q,\lambda| \, \gamma^{\mu}\frac{\rlap/ \!P - \rlap/ \!K}{-2PK} \,
\Sigma^b((P-K)^2,m_0,m_e)(\rlap/ \!P - \rlap/ \!K)  \, \frac{\rlap/ \!P - \rlap/ \!K}{-2PK}\, \rlap/ \varepsilon \, |P,\lambda \rangle  \nonumber \\
&=& \frac{ie^5}{16\pi^2} \, \Sigma^b((P-K)^2,m_0,m_e) \,  \langle Q',\lambda'| \, \gamma_{\mu} |P',\lambda'
\rangle G_{\lambda,\lambda\!'}(t') \times \nonumber \\
& &  \langle Q,\lambda| \, \gamma^{\mu}\frac{\rlap/ \!P - \rlap/ \!K}{-2PK}
\, \rlap/ \varepsilon \, |P,\lambda \rangle  \nonumber \\
&=& \frac{e^2}{16\pi^2} \, \Sigma^b((P-K)^2,m_0,m_e) \, A^{TL}\!\!\!\!\!\!_{e^{-}} \label{eq:selfen}
\end{eqnarray}

\noindent With the notation of Eq. \ref{eq:gammaparam} one finally gets:

\begin{equation}
\!\!\!\!\!\!\!\!\!\!\!\!\!\!\!\!\!\!\!\!\!\!\!\!\!\!\!\!\!\!\!\!\!\!\!\!\!\!\!\!\!\!\!\!\!\!\!\!\!\!\!\!\!\!\!A_5 + A_6 = \frac{e^2}{16\pi^2} \, \Sigma^b(\gamma,m_0,m_e)\, A^{TL}\!\!\!\!\!\!_{e^{-}} \label{eq:a5a6res}
\end{equation}

\noindent It is interesting to observe, that the IR-divergences contained in
Eq. \ref{eq:a5a6res} are exactly canceled by the $ln \frac{m_0}{m_e}$ terms
present in the expression for the amplitudes $A_7$ and $A_8$ in Eqs. 
\ref{eq:A7res} and \ref{eq:A8res}. This set of graphs is also separately
gauge invariant.

\noindent Standard Feynman techniques give $18$ one photon exchange diagrams for the process considered here (modulo vacuum polarization) as was mentioned in section \ref{sec:singBramp}. From Eqs. \ref{eq:renorm1}, \ref{eq:renorm2} it can readily be seen that all graphs with self-energy loops next to external lines don't contribute to the ${\cal O}(\alpha^2)$ calculation in the chosen on-shell renormalization scheme, since

\begin{equation}
\frac{\Sigma(P,m_0,m_e)}{\rlap/ \!P - m_e} \,\,\, \stackrel{\rlap/ \!P=m_e}{\longrightarrow} \,\,\, \frac{\partial \Sigma(P,m_0,m_e)}{\partial \rlap/ \!P} \,\,\, \stackrel{\rlap/ \!P=m_e}{\longrightarrow} \,\,\,0 \label{contact}
\end{equation}

\noindent Thus 

\begin{equation}
A_{11} \,\, = \,\, 0 \;\; , \; ... \; , \;\; A_{18} \,\, = \,\, 0 \label{A11A18}
\end{equation}

\noindent This completes the ${\cal O}({m_e}^0)$ contributions of the ${\cal O}(\alpha^2)$ single bremsstrahlung amplitudes with real hard photon emission off external legs. The next section will now deal with the much more complicated situation given for internal photon emission in amplitudes Eqs. \ref{eq:A3}, \ref{eq:A4}.

\section{The Internal Emission Amplitudes} \label{sec:intem}

\subsection{Scalar Integrals} \label{sec:intemscalint}
 
\noindent The FORM notation of scalar integrals used in the decomposition of 
the tensor integrals \ref{eq:A3}, \ref{eq:A4} differs from the notation used 
in previous sections for the obvious reason that there are now four 
propagators inside the loop. This leads to the following identifications:

\begin{eqnarray}
B_{12} &=& \! B_0({m_e}^2;m_0,m_e)    \\ \label{eq:B12}
{B_{13}\!\!\!\!\!^{\alpha}} \,  &=& \! B_0({m_e}^2 \! -\alpha;m_0,m_e) \\ \label{eq:B13al}
B_{24} &=& \! B_0(2{m_e}^2 \! +t;m_e,m_e)    \\ \label{eq:B24}
B_{34} &=& \! B_0(2{m_e}^2 \! +t';m_e,m_e)    \\ \label{eq:B34}
{C_{123}\!\!\!\!\!\!\!^{\alpha}} \,\,\, &=& \! C_0({m_e}^2,{m_0}^2,{m_e}^2 \! -\alpha;m_0,m_e,m_e) \\ \label{eq:C123al}
C_{124} &=& \! C_0({m_e}^2,2{m_e}^2 \! +t,{m_e}^2;m_0,m_e,m_e) \\ \label{eq:C124}
{C_{134}\!\!\!\!\!\!\!^{\alpha}} \,\,\, &=& \! C_0({m_e}^2 \! -\alpha,2{m_e}^2 \! +t',{m_e}^2;m_0,m_e,m_e) \\ \label{eq:C134al}
C_{234} &=& \!C_0({m_0}^2,2{m_e}^2 \! +t',2{m_e}^2 \! +t;m_e,m_e,m_e) \label{eq:C234} \\
{D_{1234}\!\!\!\!\!\!\!\!\!\!^{\alpha}} \,\,\,\,\, &=& \! D_0({m_e}\!\!^2 , \! {m_0}\!\!^2, \! 2{m_e}\!\!^2 \! +t' \! , \! {m_e}\!\!^2, \! {m_e}\!\!^2 \! -\alpha  , \! 2{m_e}\!\!^2 \! +t  ; \! m_0, \! m_e, \! m_e, \! m_e) \,, \label{eq:D1234al} 
\end{eqnarray}

\noindent where the basic $B_0, C_0 $ and $D_0$ functions are defined in section \ref{sec:scalintdef} as usual.

\subsection{The $A_3+A_4$ Calculation}
 
\noindent The numerator of Eq. \ref{eq:A3} was put into the FORM-file $a3a4$ in which the outgoing term \ref{eq:A4} was added through the crossing relation \ref{eq:cross1}. By means of algebraic manipulations, the initial output of roughly $20 000$ (!) terms could be reduced to $90$, which involved many rather complicated transformations (see appendices of Ref. \cite{thesis}). Again, the identity \ref{eq:id3} was crucial to achieve this step down to so relatively few terms as well as separating out the part proportional to $A^{TL}\!\!\!\!\!\!_{e^{-}}$. The result given by $a3a4$ reads

\begin{equation}
A_3 + A_4 = {ie^5\over 16\pi^2} G_{\lambda, \lambda'}(t')
\left(  F_{0} {\cal I}_0 +  F_{1} {\cal I}_1 +
 F_{2} {\cal I}_2 \right) \;, \label{eq:a3a4res}
\end{equation}
\noindent where
\begin{eqnarray}
{\cal I}_0 &=& 2\sqrt{2}\rho \langle P', Q'\rangle_{-\rho}
{\left(\langle h, h' \rangle_\rho\right)^2\over
 \langle P,K\rangle_{\rho} \langle Q,K \rangle_{\rho}}\, \label{eq:I0def} \\
{\cal I}_1 &=& 2\sqrt{2}\lambda
{\langle {\widehat h},K\rangle_{-\rho}\over
 \langle {\widehat h}, K \rangle_{\rho}}
{\langle Q',\lambda'| \not \!h \,| P', \lambda' \rangle\over
 \langle P,Q \rangle_{-\rho}}\  \label{eq:I1def} \\
{\cal I}_2 &=& 2\sqrt{2}\lambda
{\langle {\widehat h},K\rangle_{-\rho}\over
 \langle {\widehat h}, K \rangle_{\rho}}
{\langle Q',\lambda'| \not \!{\widehat h} \, | P',\lambda' \rangle\over
 \langle P,Q \rangle_{-\rho}}\  \label{eq:I2def}
\end{eqnarray}
\noindent The function ${\cal I}_0$ is proportional to the electron-line tree level amplitude:
\begin{equation}
A^{TL}\!\!\!\!\!\!_{e^{-}} = i e^3 G_{\lambda, \lambda'}(t')\ {\cal I}_0\ 
\end{equation}
\noindent The form factors are
\begin{eqnarray}\lefteqn{{F}_0(\rho=\lambda)= 
- 2(tC_{124} - t'C^{-\beta}_{134})-(B^{-\beta}_{13}-B_{34})3\beta(t-\alpha)^{-1}}\nonumber\\
&-&\{tC_{124}\; - {\alpha}C^{\alpha}_{123}\; - (t + \beta)C^{\alpha}_{134}\; + (
\alpha - \beta)C_{234}\; + {\alpha}tD^{\alpha}_{1234}\;\} \nonumber\\ & & \qquad
\qquad\times\ t\beta^{-1}(\alpha - \beta)(t - \alpha)^{-1} \nonumber\\
&+&\{tC_{124}\; + {\beta}C^{-\beta}_{123}\; - (t - \alpha)C^{-\beta}_{134}\; + (
\alpha - \beta)C_{234}\; - {\beta}tD^{-\beta}_{1234}\;\}  \nonumber\\
&+& (B^{\alpha}_{13}-B_{34})\;\alpha(t-\alpha)^{-1}\;\{1 - 3t(t + \beta)^{-1} \} \nonumber\\
&+& (B_{24}-B_{34})\;2t\alpha(\alpha - \beta)^{-1}(t - \alpha)^{-1} \; + \; (B_{12} - B^{-\beta}_{13}) 2t(t-\alpha)^{-1}
 \nonumber\\
\end{eqnarray}\begin{eqnarray}\lefteqn{{F}_1(\rho=\lambda)= 2t(\alpha - \beta)^{-1} }\nonumber\\
&+&\{tC_{124}\; - {\alpha}C^{\alpha}_{123}\; - (t + \beta)C^{\alpha}_{134}\; + (
\alpha - \beta)C_{234}\; + {\alpha}tD^{\alpha}_{1234}\;\} \nonumber\\ & & \qquad
\qquad\times\{t{t^{\prime}}\beta^{-2}(t - \alpha)^{-1}(t - \beta) +{\mbox{${1\over2}$}}\delta_{\rho,1} \} \nonumber\\
&-&\{tC_{124}\; + {\beta}C^{-\beta}_{123}\; - (t - \alpha)C^{-\beta}_{134}\; + (
\alpha - \beta)C_{234}\; - {\beta}tD^{-\beta}_{1234}\;\} \nonumber\\ & & \qquad
\qquad\times\{{\mbox{${1\over2}$}}\alpha^{-1}\beta\delta_{\rho,-1} \} \nonumber\\
&+&(B^{-\beta}_{13}-B_{12})\;2t(t - \alpha)^{-1} \nonumber\\
&+&(B^{\alpha}_{13}-B_{34})\;tt'(t - \alpha)^{-1}\;\{2\beta^{-1} -3(t + \beta)^{-1} \} \nonumber\\
&+&2(B_{24}-B_{34})\;tt'(\alpha - \beta)^{-1}\{(\alpha - \beta)^{-1}-t\beta^{-1}
(t - \alpha)^{-1}  \} \nonumber\\
\end{eqnarray}\begin{eqnarray}\lefteqn{{F}_2(\rho=\lambda)= -2t(\alpha
- \beta)^{-1} + t(t + \beta)^{-1} }\nonumber\\
&-&\{tC_{124}\; - {\alpha}C^{\alpha}_{123}\; - (t + \beta)C^{\alpha}_{134}\; + (
\alpha - \beta)C_{234}\; + {\alpha}tD^{\alpha}_{1234}\;\} \nonumber\\ & & \qquad
\qquad\times\{t{t^{\prime}}\beta^{-2} + {\mbox{${1\over2}$}}\alpha\beta^{-1}\delta_{\rho,1} \} \nonumber\\
&+&\{tC_{124}\; + {\beta}C^{-\beta}_{123}\; - (t - \alpha)C^{-\beta}_{134}\; + (
\alpha - \beta)C_{234}\; - {\beta}tD^{-\beta}_{1234}\;\} \nonumber\\ & & \qquad\qquad\times\{{\mbox{${1\over2}$}}\delta_{\rho,-1} \} \nonumber\\
&+&(B_{34}-B^{\alpha}_{13})\;tt'(t+\beta)^{-1}\;\{2\beta^{-1} + (t + \beta)^{-1}
 \} \nonumber\\
&+&2(B_{24}-B_{34})\;tt'(\alpha-\beta)^{-1}\;\{\beta^{-1} -(\alpha - \beta)^{-1}
 \} \nonumber\\
\end{eqnarray}

\noindent The opposite helicity cases may be obtained from the above results using the substitutions (for $i=0,1,2$):

\begin{equation}
{F}_i(\rho=-\lambda, \alpha,\beta) =
{F}_i(\rho=\lambda, -\beta, -\alpha)\ 
\end{equation}

\noindent While Eq. \ref{eq:a3a4res} still does not look very transparent, 
several observations may help to clarify the overall picture. Since the
amplitudes \ref{eq:A3} and \ref{eq:A4} are not UV-divergent one should
evidently expect this feature for the result \ref{eq:a3a4res}, where all
$B_0$ terms possess such an infinite contribution (see section 
\ref{sec:scalintdef}). A simple check in the FORM routine $a3a4$, which  
replaced each $B_{ij}$ by $B_{ij}+div$, proved that all divergences cancel
properly. This procedure also insures the scale independence of Eq. 
\ref{eq:a3a4res}. Numerical analysis revealed that in the 
case of amplitude $A_3$, 
for instance, only $C_{124}$ and ${D_{1234}\!\!\!\!\!\!\!\!\!\!^{\alpha}}
 \;\;\;$ 
have IR-divergent terms, regulated by the virtual photon mass $m_0$. By
closer examining the four denominators in \ref{eq:A3}, it becomes clear that
, except for $K^0=0$, only propagators $1, 2 \; \& \; 4$ 
taken together and with no
loop momentum left in the numerator can lead to a logarithmic divergence in
$m_0$, confirming the numerical findings. From Ref. \cite{dittm93} and for 
$\frac{m_e\!^2}{-t} \ll 1$ one has:

\begin{equation}
C_{124} \;\; = \;\; \frac{1}{t} \left[ \frac{1}{2} ln^2 \frac{m_e\!^2}{-t} \; 
+ \; ln \frac{m_e\!^2}{-t} ln \frac{m_0\!^2}{m_e\!^2} \; - \; \frac{\pi^2}{6} 
\right] \;, \label{eq:C124res}
\end{equation}

\noindent and for $m_0 \rightarrow 0$:

\begin{equation}
{D_{1234}\!\!\!\!\!\!\!\!\!\!^{\alpha}} \;\;\;\; \sim \; - \; 
\frac{1}{\alpha t}
ln \frac{m_e\!^2}{-t} ln \frac{m_0\!^2}{m_e\!^2} \label{eq:Dlimit}
\end{equation}

\noindent Eqs. \ref{eq:C124res} and \ref{eq:Dlimit} corroborate that the only
effective IR-divergent term in \ref{eq:a3a4res} is 
$-2tC_{124} \frac{ie^5}{16\pi^2} {\cal I}_0$, which 
coincides with the prediction of the IR-theory expounded on in chapter 
\ref{sec:YFS}. It should also be noted that the form factors $F_1$ and $F_2$
are normalized in a slightly different way than the output printed in
Ref. \cite{thesis}. For comparison, the form factors 
here should be multiplied by
$\frac{2}{t\beta}$ for $\rho = \lambda$. $F_1$ contains
a contribution proportional to $(B^{-\beta}_{13}-B_{12})$ which will turn
out to cancel the corresponding contribution from amplitude \ref{eq:A7res}
and will thus make $F_1$ and $F_2$ almost symmetric in form. This feature
will help concentrate the numerically relevant leading and subleading 
logarithmic terms in the part proportional to the tree level amplitude, as
the results of this chapter will now be summarized in chapter \ref{sec:exres}.

\chapter{The Exact Differential ${\cal O}(\alpha^2)$ Result} \label{sec:exres}

\noindent In order to have a complete differential result it is
necessary to express all amplitudes with the same notation of scalar 
integrals. Using the definitions of section \ref{sec:intemscalint} and the
expressions given in Eqs. \ref{eq:a1a2res}, \ref{eq:a9a10res}, \ref{eq:A7res},
\ref{eq:A8res} and \ref{eq:a5a6res} one gets the following for 
$\rho = -\lambda$:

\begin{eqnarray}
&& A_1+A_2+A_5+A_6+A_7+A_8+A_9+A_{10} \nonumber \\
&=& \frac{e^2}{16\pi^2} \left\{ -8 - \frac{\beta}{t+\beta} + B_{12} 
\frac{4t + 6\beta}{t+\beta} + B^{\alpha}_{13} \frac{2t-3\alpha}{t'+\alpha}
\right. \; - \nonumber \\
&& \left.  B_{34} \frac{6t'+3\alpha}{t'+\alpha} -  8 m_e\!^2 C^0_{123} - 2t' C^{\alpha}_{134} - 2t' C^0_{134} \right\} 
A^{TL}\!\!\!\!\!\!_{e^{-}} \nonumber \\
&& - \; \frac{ie^5}{16\pi^2} G_{\lambda, \lambda'}(t') \left[2B^{\alpha}_{13} 
-2B_{12} + 1 \right] \frac{t}{t+\beta} \; {\cal I}_1 \;, \label{eq:a125678910res}
\end{eqnarray}

\noindent and thus for $\rho=\lambda$:

\begin{eqnarray}
&& A_1+A_2+A_5+A_6+A_7+A_8+A_9+A_{10} \nonumber \\
&=& \frac{e^2}{16\pi^2} \left\{ -8 + \frac{\alpha}{t-\alpha} + B_{12} 
\frac{4t - 6\alpha}{t-\alpha} + B^{-\beta}_{13} \frac{2t+3\beta}{t'-\beta}
\right. \; - \nonumber \\
&& \left. B_{34} \frac{6t'-3\beta}{t'-\beta} -  8 m_e\!^2 C^0_{123} - 2t' C^{-\beta}_{134} - 2t' C^0_{134} \right\} 
A^{TL}\!\!\!\!\!\!_{e^{-}} \nonumber \\
&& - \; \frac{ie^5}{16\pi^2} G_{\lambda, \lambda'}(t') \left[2B^{-\beta}_{13} 
-2B_{12} + 1 \right] \frac{t}{t-\alpha} \; {\cal I}_1 \label{eq:b125678910res}
\end{eqnarray}

\noindent The exact ${\cal O}(\alpha^2)$ single bremsstrahlung result for 
electron-line emission and $\rho = \lambda$ is of course given by adding all
amplitudes \ref{eq:A1} + ... + \ref{eq:A10} and with Eqs. \ref{eq:a3a4res}
and \ref{eq:b125678910res} reads for $\rho=\lambda$:

\begin{equation}
A_{e^-} = {ie^5\over 16\pi^2} G_{\lambda, \lambda'}(t')
\left( {\cal F}_{0} {\cal I}_0 + {\cal F}_{1} {\cal I}_1 +
{\cal F}_{2} {\cal I}_2 \right) \;, \label{eq:exres}
\end{equation}

\noindent where the notation of Eqs. \ref{eq:I0def}, \ref{eq:I1def} and \ref{eq:I2def} has been used.
The form factors are

\begin{eqnarray}\lefteqn{{\cal{F}}_0(\rho=\lambda)= -8 + \alpha(t - \alpha)^{-1}
- 8m_e\!^2 C^0_{123} - 2(tC_{124} + t'C_{134}^0)}\nonumber\\
&-&\{tC_{124}\; - {\alpha}C^{\alpha}_{123}\; - (t + \beta)C^{\alpha}_{134}\; + (
\alpha - \beta)C_{234}\; + {\alpha}tD^{\alpha}_{1234}\;\} \nonumber\\ & & \qquad
\qquad\times\ t\beta^{-1}(\alpha - \beta)(t - \alpha)^{-1} \nonumber\\
&+&\{tC_{124}\; + {\beta}C^{-\beta}_{123}\; - (t - \alpha)C^{-\beta}_{134}\; + (
\alpha - \beta)C_{234}\; - {\beta}tD^{-\beta}_{1234}\;\}  \nonumber\\
&+&6B_{12}\ +\ (B^{\alpha}_{13}-B_{34})\;\alpha(t-\alpha)^{-1}\;\{1 - 3t(t + \beta)^{-1}\} \nonumber\\
&-&6B_{34}\ +\ (B_{24}-B_{34})\;\{2t\alpha(\alpha - \beta)^{-1}(t - \alpha)^{-1}
 \} \nonumber\\
\end{eqnarray}\begin{eqnarray}\lefteqn{{\cal{F}}_1(\rho=\lambda)= 2t(\alpha - \beta)^{-1} - t(t - \alpha)^{-1}}\nonumber\\
&+&\{tC_{124}\; - {\alpha}C^{\alpha}_{123}\; - (t + \beta)C^{\alpha}_{134}\; + (
\alpha - \beta)C_{234}\; + {\alpha}tD^{\alpha}_{1234}\;\} \nonumber\\ & & \qquad
\qquad\times\{t{t^{\prime}}\beta^{-2}(t - \alpha)^{-1}(t - \beta) +{\mbox{${1\over2}$}}\delta_{\rho,1} \} \nonumber\\
&-&\{tC_{124}\; + {\beta}C^{-\beta}_{123}\; - (t - \alpha)C^{-\beta}_{134}\; + (
\alpha - \beta)C_{234}\; - {\beta}tD^{-\beta}_{1234}\;\} \nonumber\\ & & \qquad
\qquad\times\{{\mbox{${1\over2}$}}\alpha^{-1}\beta\delta_{\rho,-1} \} \nonumber\\
&+&(B^{\alpha}_{13}-B_{34})\;tt'(t - \alpha)^{-1}\;\{2\beta^{-1} -3(t + \beta)^{-1} \} \nonumber\\
&+&2(B_{24}-B_{34})\;tt'(\alpha - \beta)^{-1}\{(\alpha - \beta)^{-1}-t\beta^{-1}
(t - \alpha)^{-1}  \} \nonumber\\
\end{eqnarray}\begin{eqnarray}\lefteqn{{\cal{F}}_2(\rho=\lambda)= -2t(\alpha
- \beta)^{-1} + t(t + \beta)^{-1} }\nonumber\\
&-&\{tC_{124}\; - {\alpha}C^{\alpha}_{123}\; - (t + \beta)C^{\alpha}_{134}\; + (
\alpha - \beta)C_{234}\; + {\alpha}tD^{\alpha}_{1234}\;\} \nonumber\\ & & \qquad
\qquad\times\{t{t^{\prime}}\beta^{-2} + {\mbox{${1\over2}$}}\alpha\beta^{-1}\delta_{\rho,1} \} \nonumber\\
&+&\{tC_{124}\; + {\beta}C^{-\beta}_{123}\; - (t - \alpha)C^{-\beta}_{134}\; + (
\alpha - \beta)C_{234}\; - {\beta}tD^{-\beta}_{1234}\;\} \nonumber\\ & & \qquad\qquad\times\{{\mbox{${1\over2}$}}\delta_{\rho,-1} \} \nonumber\\
&+&(B_{34}-B^{\alpha}_{13})\;tt'(t+\beta)^{-1}\;\{2\beta^{-1} + (t + \beta)^{-1}
 \} \nonumber\\
&+&2(B_{24}-B_{34})\;tt'(\alpha-\beta)^{-1}\;\{\beta^{-1} -(\alpha - \beta)^{-1}
 \} \nonumber\\
\end{eqnarray}

\noindent The opposite helicity cases may again be obtained from the above results using the substitutions (for $i=0,1,2$):

\begin{equation}
{\cal F}_i(\rho=-\lambda, \alpha,\beta) =
{\cal F}_i(\rho=\lambda, -\beta, -\alpha)\ 
\end{equation}

\begin{figure}
\epsfig{file=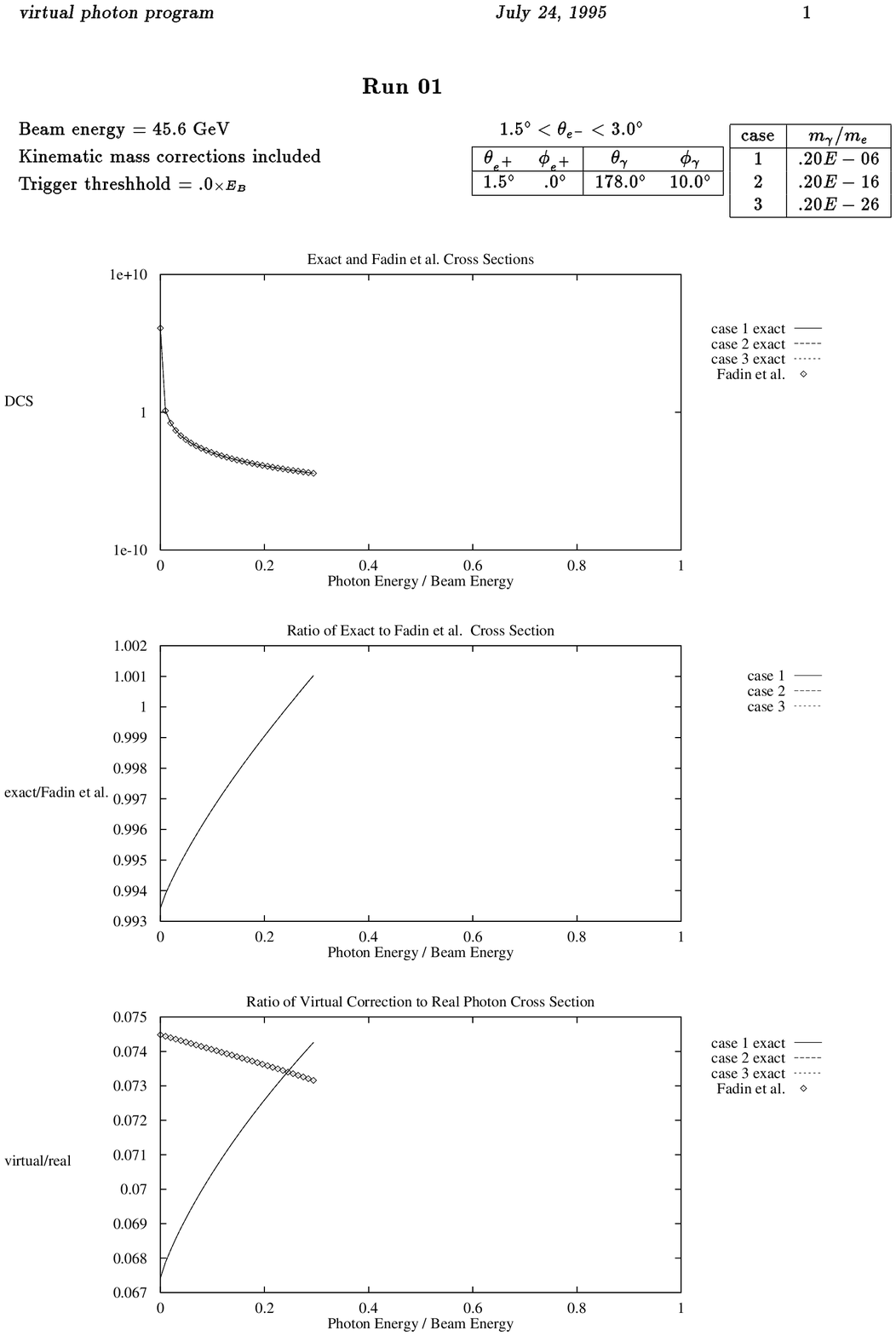,height=17cm}
\caption{A comparison of the exact differential expression \protect\ref{eq:exres} to the result given by Fadin et al. \protect\cite{kuraev94}. 
The graph on top displays $|{\cal M}^{{\cal O}(\alpha+\alpha^2)}|^2
- 2 \alpha [Re(B(t)) + Re(B(t'))]|{\cal M}^{{\cal O}(\alpha)}|^2$ (see chapter \protect\ref{sec:YFS}), 
i.e. there is no IR-dependence ($m_0$ or $E_{cut}$) left over. ${\cal M}$
denotes the invariant matrix elements for the respective cases. 
The second figure 
shows the ratio of the two calculations explicitly over the range of 
available photon energies and the bottom graph gives the size of the 
${\cal O}(\alpha^2)$ virtual corrections to the differential 
${\cal O}(\alpha)$ single hard bremsstrahlung cross section.}
\label{fig:run1}
\end{figure}

\begin{figure}
\epsfig{file=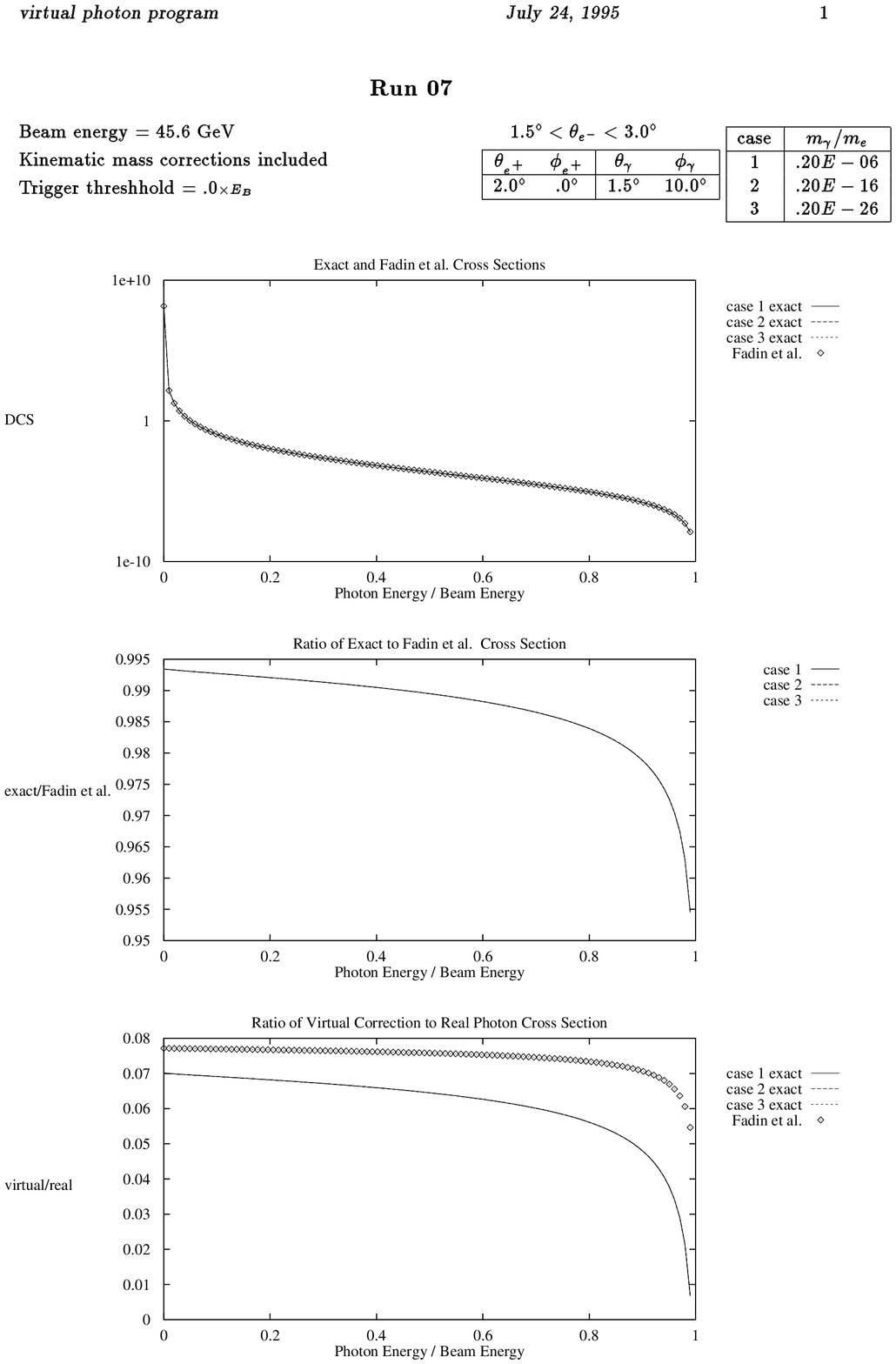,height=17cm}
\caption{This figure is analogous to Fig \ref{fig:run1} but shows the same comparison for a different set of photon and positron angles. While run1 has the bremsstrahlung photon close to the incoming electron direction, this graph has the emitted photon along the incoming positron line. Although the respective lines are dominating in each case, both figures are using the contribution from the complete two line t-channel expression.}
\label{fig:run7}
\end{figure}

\noindent In Figs. \ref{fig:run1} and \ref{fig:run7} the differential
result \ref{eq:exres}
is compared to a recent calculation by Fadin et al. \cite{kuraev94} after the 
IR-divergent terms were properly removed (see chapter \ref{sec:YFS}).

\noindent The agreement between these two differential results is roughly
within $5\%$ over the range of available real photon energies. This should be 
compared to the overall size of the second order correction to the Born
process, which is about $3.5 pm$ \cite{Been93}. The soft limit,
however, disagrees, which will become more important for the integrated
cross section since it is the regime with the highest emission probability
\cite{Piet94}. As can also be seen in both figures, there is no virtual photon-mass
dependence left. The one graph that is shown is actually three identical 
graphs for three different values of $m_0$ (compare with figure legend).
The general treatment of the expected IR-behaviour is discussed in chapter
\ref{sec:YFS} in the context of Yennie, Frautschi and Suura theory \cite{yfs61}.
The expression is explicitly scale independent, as is expected of every
physically meaningful result. This holds not only for \ref{eq:exres} but also
for all equations that have contributed to the final answer.
It is also worth mentioning that the by far largest part
of the answer \ref{eq:exres} is concentrated in only $18$ terms, namely
those contained in ${\cal F}_0$. Numerical tests showed that on the scale of
Figs. \ref{fig:run1} and \ref{fig:run7} no noticeable change could be observed.
Since these represent only one dimensional cuts through a multidimensional
phase space, this aspect needs to be revisited in the context of 
Monte Carlo (MC)
phase space integration.

\noindent Eq. \ref{eq:exres} is the central result of the presented work;
it describes the fully differential ${\cal O}(\alpha^2)$ single bremsstrahlung
corrections to low
angle Bhabha scattering, which is essential for any MC-implementation 
\cite{JadRiWa92}. It is thus of primary importance for the theoretical luminosity
determination below the $1pm$ precision level. In order to convey
a more in depth understanding about the validity of the presented work,
part $II$ of this thesis will mostly be devoted to demonstrate that 
Eq. \ref{eq:exres} satisfies all theoretical expectations and is in good
agreement with leading log calculations. Also Monte Carlo results for the
fully integrated cross section will be presented for a SICAL-type
acceptance \cite{Been93,JadRiWa95}.

\part{Consistency Checks \& MC Results}

\chapter{Yennie, Frautschi \& Suura Theory} \label{sec:YFS}

\noindent Although quantum field theories with massless gauge bosons display
both short- (UV) and long-distance (IR) divergences in certain types
of graphs, it has long been known that the nature of the two infinities have 
very different origins \cite{kaku,greschwit1}. While the UV-divergences are a manifestation 
of our ignorance about small-scale physics and cannot be avoided in theories
that assume point-like elementary particles, the genuine infinite wavelength 
divergences should never
be "real", i.e. should cancel out of the final result for any cross section.  
The cancelation of the first order IR-divergences in QED was first proven by 
Bloch and Nordsieck \cite{bloch37} as early as 1937. It took, however, almost
another quarter of a century until the IR-problem was understood 
comprehensively to all orders in the fundamental work by Yennie, Frautschi 
and Suura (YFS) \cite{yfs61}. Since the YFS-theory gives a well defined 
IR-limit for the amplitude \ref{eq:a3a4res}, it is necessary at this point to
give a brief review of the IR-theory. $\alpha$ in this chapter denotes the 
fine-structure constant.

\section{The YFS $B$ \& $\widetilde{B}$ Functions} \label{sec:YFSB}

\noindent Consider the expansion of the full connected amplitude for s-channel Bhabha scattering at $\sqrt{s} = M_Z$ in terms of the number of virtual photon loops. One can then write \cite{yfs61}

\begin{equation}
M(P,P') \; = \; \sum_{n=0}^{\infty} M_n(P,P') \;, \label{eq:YFSexpan} 
\end{equation}

\noindent where $M_n(P,P')$ is the contribution of all n virtual $\gamma$-loop graphs to $M(P,P')$. The result obtained by Yennie, Frautschi and Suura is that

\begin{equation}
M_n(P,P') \; = \; \sum_{r=0}^{n} m_{n-r}(P,P') \frac{(\alpha B)^r}{r \, !} \;,
\end{equation}

\noindent where the $m_j(P,P')$ do not have virtual IR-divergences and are of order $\alpha^j$ relative to $M_0 = m_0$. The virtual infrared function $B(s)$ is such that

\begin{eqnarray}
B(s) \! &=& \!\! \frac{i}{(2\pi)^3} \int \frac{d^4k}{k^2 - {m_0}\!^2 + i\epsilon}\! \left[ \frac{-2P'_{\mu} - k_{\mu}}{-2P'k - k^2 - i \epsilon} - \frac{2P_{\mu} - k_{\mu}}{2Pk-k^2 - i \epsilon} \right ]^2 \;,\nonumber \\[0.2cm]
Re \; B(s) \! &=& \!\! \frac{-1}{2 \pi} \left[ ln \frac{s}{{m_e}\!^2} \! \left(\! 2 ln \frac{m_e}{m_0} + \frac{1}{2} ln \frac{s}{{m_e}\!^2} - \frac{1}{2} \right) - 2 ln \frac{m_e}{m_0} + 1 - \frac{2 \pi^2}{3} \right] \label{eq:Bdef}
\end{eqnarray}

\noindent Hence

\begin{equation}
M(P,P') \; = \; e^{\alpha B} \sum_{n=0}^{\infty} m_n(P,P') \label{eq:virtexp}
\end{equation}

\noindent This is the famous exponentiation of virtual infrared divergences of the YFS program. To complete the review of the YFS-theory, consider next the differential cross section for the process $e^+ + e^- \rightarrow n(\gamma) + X$ , where $n(\gamma)$ represents the emission of $n$ real photons with four-momenta $k_1, ... , k_n$. For a given value of $n$, this differential cross section is 

\begin{equation}
d \sigma = e^{2 \alpha Re \, B} \frac{1}{n \,!} \int \prod_{j=1}^{n} \frac{dE_X d^3k_j}{\sqrt{{k_j}\!^2 + {m_0}\!^2}} \, \delta \! \left( \sqrt{s} - E_X - \sum_{i=1}^n {k_i}\!^0 \right) \! \left[ \sum_{l=0}^{\infty} {m_l}\!^{(n)} \right]^2 \;, \label{eq:YFScs1}
\end{equation}

\noindent where ${m_l}\!^{(n)}$ is a special case of $m_j$ in \ref{eq:virtexp} in which X involves n real photons. The second theorem of Yennie, Frautschi and Suura is that

\begin{eqnarray}
\left[ \sum_{l=0}^{\infty} {m_l}\!^{(n)}\right]^2 \!\!\!\!\! &=& \!\!\! \widetilde{S}(k_1) ... \widetilde{S}(k_n) \widetilde{\beta}_0 + \sum_{i=1}^{n} \widetilde{S}(k_1) ... \widetilde{S}(k_{i-1}) \widetilde{S}(k_{i+1}) ... \widetilde{S}(k_n) \widetilde{\beta}_1(k_i) + ... \nonumber \\
\!\!\!\!\! & & \!\!\! + \sum_{i=1}^{n} \widetilde{S}(k_i) \widetilde{\beta}_{n-1}(k_1,...,k_{i-1},k_{i+1},...,k_n) + \widetilde{\beta}_n (k_1,...,k_n) \;, \label{eq:YFSbeta}
\end{eqnarray}

\noindent where $\widetilde{\beta}_j$ is IR-divergence free and is of order $\alpha^j$ relative to $\widetilde{\beta}_0$. 
The real infrared divergence function $\widetilde{S}(k)$ is given by

\begin{equation}
\widetilde{S}(k) \; = \; - \; \frac{\alpha}{4\pi^2} \left[ \frac{P_{\mu}}{kP} - \frac{P'_{\mu}}{kP'} \right]^2 \label{eq:YFSsdef}
\end{equation}

\noindent It follows that the cross section for the emission of an arbitrary number of real photons can be written as

\begin{eqnarray}
d \sigma  &=&  e^{2\alpha (Re \, B + \widetilde{B})} \frac{1}{2\pi} \int_{-\infty}^{\infty} \!\!\! dy \;  e^{iy(\sqrt{s}-E_X) + D} \;\;\;\;\; \times \nonumber \\  & & \left[ \widetilde{\beta}_0 + \sum_{n=1}^{\infty} \frac{1}{n \, !} \int \prod_{j=1}^n \frac{d^3 k_j}{{k_j}\!^0} e^{-iy{k_j}\!^0} \widetilde{\beta}_n \right] d E_X \;, \label{eq:YFScs}
\end{eqnarray}

\noindent with $D$ and $\widetilde{B}$ defined as follows:

\begin{equation}
\!\!\!\!\!\!\!\!\!\!\!\!\!\!\!\!\!\!\!\!\!\!\!\!\!\!\!\!\!\!\!{D \;\;\; = \;\; \int^{k^0 \leq E_{cut}} \frac{d^3 k}{k^0} ( e^{-iyk^0} -1) \widetilde{S}} \;, \label{eq:YFSDdef}
\end{equation}

\noindent and

\begin{eqnarray}
\widetilde{B}(s,E_{cut}) &=& \int^{k^0 \leq  E_{cut}} \frac{d^3 k}{\sqrt{k^2 + {m_0}\!^2}} \frac{\widetilde{S}(k)}{2\alpha} \nonumber \\[0.2cm]
&=& \frac{1}{2\pi} \left[ ln \frac{s}{{m_e}\!^2} \left( 2 ln \frac{m_e}{m_0} 
+ \frac{1}{2} ln \frac{s}{{m_e}\!^2} - ln \frac{EE'}{{E_{cut}\!\!\!\!\!\!^2}} \; \right) \right. \nonumber \\ 
& & \;\;\;\;\;\;\; \left. - 2 ln \frac{m_e}{m_0} + ln \frac{EE'}{{E_{cut}\!\!\!\!\!\!^2}} - \frac{\pi^2}{3} \right] \label{eq:Btildedef}
\end{eqnarray}

\noindent From the explicit forms in Eqs. \ref{eq:Bdef} and \ref{eq:Btildedef} it can be verified that $Re \, B + \widetilde{B}$ is free of IR-divergences so that $d \sigma$ in \ref{eq:YFScs} is indeed a physically meaningful quantity and exhibits the cancelation of infrared divergences to all orders in $\alpha$. 
It should be mentioned that the infrared functions given above are expressed in the s-channel and that the t-channel expressions are derived from \ref{eq:Bdef} and \ref{eq:Btildedef} with the analytical continuation technique and read \cite{JadWard88}:

\begin{eqnarray}
\!\!\!\!\!\!\!\!\!\!\!Re \; B(t) &=& \!\!\!\!\! \frac{-1}{2 \pi}\! \left[ ln \frac{-t}{{m_e}\!^2} \! \left(\! 2 ln \frac{m_e}{m_0} + \frac{1}{2} ln \frac{-t}{{m_e}\!^2} - \frac{1}{2} \right) - 2 ln \frac{m_e}{m_0} + 1 - \frac{ \pi^2}{6} \right] \label{eq:Btres} \\[0.2cm]
\!\!\!\!\!\!\!\!\!\!\!\widetilde{B}(t,E_{cut}) \!\!\!\!\! &=& \!\!\!\!\!
\frac{1}{2\pi} \! \left[ ln \frac{-t}{{m_e}\!^2} \left( 2 ln \frac{m_e}{m_0} + \frac{1}{2} ln \frac{-t}{{m_e}\!^2} - ln \frac{E_PE_Q}{{E_{cut}\!\!\!\!\!\!^2}} \right) \right. \nonumber \\ 
& &  \;\;\;\;  \left. - 2 ln \frac{m_e}{m_0} + ln \frac{E_PE_Q}{{E_{cut}\!\!\!\!\!\!^2}} - \frac{\pi^2}{6} \right] \label{eq:Bttildedef}
\end{eqnarray}

\begin{figure}
\epsfig{file=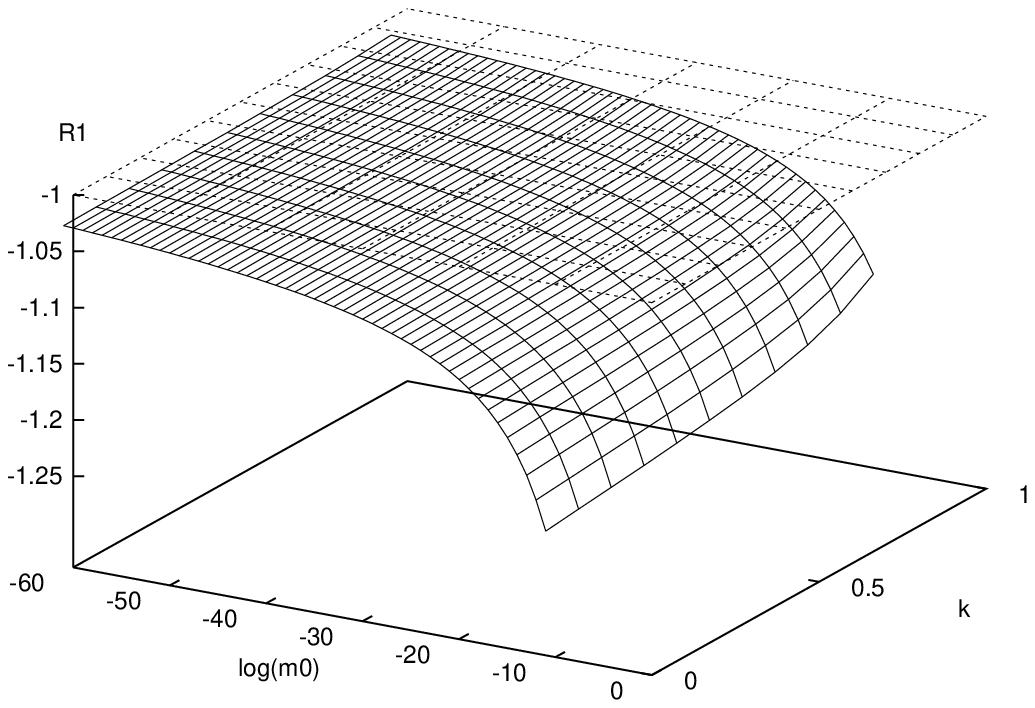,width=12cm}
\epsfig{file=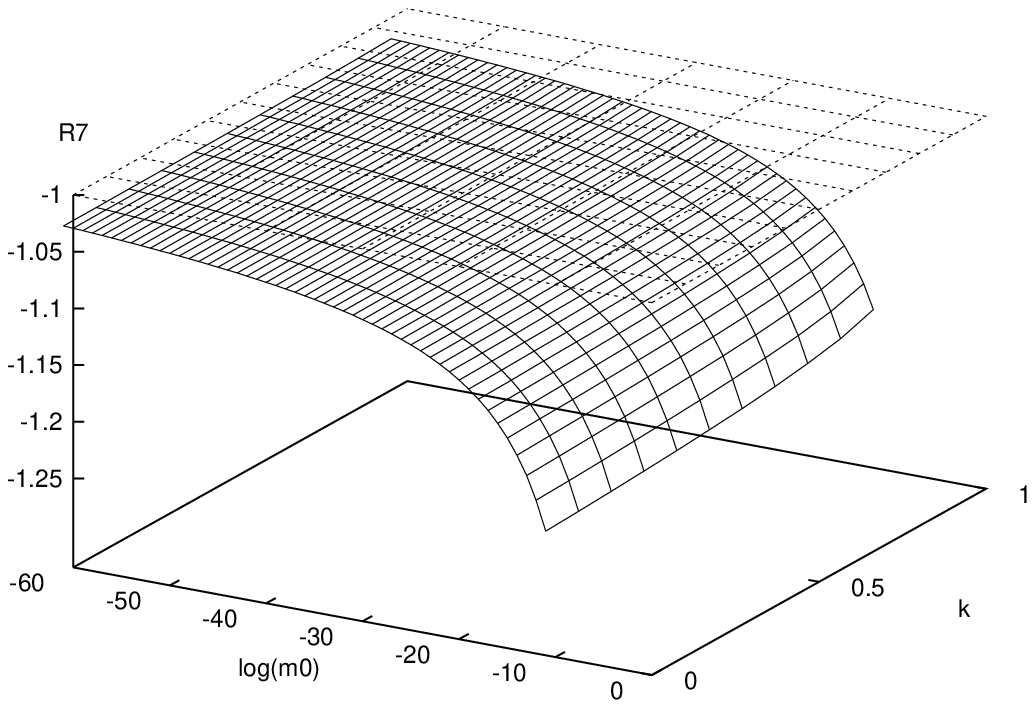,width=12cm}
 \caption{For the two separate runs, specified in Figs. \ref{fig:run1}, \ref{fig:run7} as $1$(R1) and $7$(R7), the ratios of the squared invariant matrix 
elements resulting from 
the exact differential expression \ref{eq:exres} versus 
$2 \alpha (\widetilde{B}(t,E_{cut})+\widetilde{B}
(t',E_{cut})) |{\cal M}^{{\cal O}(\alpha)}|^2$ are plotted. The limit $m_0 \longrightarrow 0$ should be $k=\frac{E_K}{E_{Beam}}$ independent and the ratios shown should approach $-1$ since the real YFS-functions 
are used. Both constraints are shown to hold for the result \ref{eq:exres}.
The finite gap to $-1$ seen in both cases is of order $\frac{1}{log(m_0)}$.}
 \label{fig:YFS}
\end{figure}

\noindent It is with the real YFS-IR function \ref{eq:Bttildedef} that the infrared limit of the internal emission result \ref{eq:a3a4res} was tested. Numerical results for two representative kinematical cases are shown in Figs. \ref{fig:YFS}. As is explained in the figure captions, the result for the amplitudes $A_1-A_{10}$ is shown to have the right IR-terms predicted by the YFS program.

\chapter{Gauge Variations} \label{sec:gaugevar}

\noindent Ever since t'Hooft was able to prove the renormalizability of spontaneously broken gauge field theories \cite{tHooft71}, gauge invariance has become the dominant principle in modern particle physics. While it was long known that QED has gauge symmetry \cite{bjdrell1,feynm1}, the weak and also the strong (QCD) nuclear force are now formulated through gauge invariance principles \cite{kaku,pokorski}. Many of the consequences of gauge symmetries such as the Ward-Takahashi identities can be used as cross checks of results obtained from related amplitudes. One such cross check in the present calculation is a result of the gauge invariance of a 
subset of amplitudes and will be expounded on in the following sections. 

\section{Gauge Invariance} \label{sec:gaugeinv}

\noindent All physically observable quantities in the Standard Model must naturally be gauge invariant, however, also unobservable parts of those quantities can already display this symmetry. It can be shown that the amplitudes $\{ A_1 + A_2 + A_3 + A_4 \}$, $\{ A_5 + A_6 + A_7 + A_8 \}$ and $\{ A_9 + A_{10} \}$ are all separately gauge invariant. This will now be shown explicitly for the first bracket of amplitudes. Gauge invariance demands that longitudinal photons decouple from the amplitudes \cite{feynm1}, thus by replacing $\rlap/ \varepsilon \rightarrow \rlap/ \!K$ one gets:

\vspace{-0.40cm}
\begin{eqnarray}
A_1\!\!\!^{long} &=& \frac{e^5}{(2\pi)^4} \langle Q',\lambda'| \, \gamma_{\mu} \, |P',\lambda' \rangle G_{\lambda,\lambda\!'}(t') \times \nonumber \\
& & \int \frac{\mu^{\epsilon}d^nl}{-2PK} \frac{\langle Q,\lambda| \, \gamma_{\nu}(\not l +
  \rlap/ \!Q) \gamma^{\mu} (\not l + \rlap/ \!P - \rlap/ \!K) \gamma^{\nu}(\rlap/ \!P - \rlap/ \!K) \, \rlap/ \!K \, |P,\lambda \rangle}
  {(l^2 - {m_0}^2)((l+P-K)^2 - {m_e}^2)((l+Q)^2 - {m_e}^2) + i\epsilon} \nonumber  \\
& & \label{eq:A1l} \\
A_2\!\!\!^{long} &=& \frac{e^5}{(2\pi)^4}  \langle Q',\lambda'| \, \gamma_{\mu} \, |P',\lambda' \rangle G_{
\lambda,\lambda\!'}(t') \times \nonumber \\
& & \int \frac{\mu^{\epsilon}d^nl}{2QK} \frac{ \langle Q,\lambda| \,\rlap/ \!K \, (\rlap/ \!Q + \rlap/ \!K) \gamma_{\nu}(\not l + \rlap/ \!Q + \rlap/ \!K) \gamma^{\mu} (\not l + \rlap/ \!P) \gamma^{\nu} \, |P,\lambda \rangle}{(l^2 - {m_0}^2)((l+P)^2 - {m_e}^2)((l+Q+K)^2 - {m_e}^2) + i\epsilon} \nonumber  \\
& & \label{eq:A2l} \\
A_3\!\!\!^{long} &=& \frac{e^5}{(2\pi)^4}   \langle Q',\lambda'| \, \gamma_{\mu} \, |P',\lambda' \rangle G_{
\lambda,\lambda\!'}(t') \times \nonumber \\
& &\!\!\!\!\!\!\!\!\!\!\!\!\!\!\!\!\!\!\!\! \int \frac{d^4l \langle Q,\lambda| \, \gamma_{\nu}(\not l +
  \rlap/ \!Q) \gamma^{\mu} (\not l + \rlap/ \!P - \rlap/ \!K) \, \rlap/ \!K \, (\not l + \rlap/ \!P) \gamma^{\nu} \, |P,\lambda \rangle}{(l^2 - {m_0}^2)((l+P)^2 - {m_e}^2)((l+P-K)^2 - {m_e}^2)((l+Q)^2 - {m_e}^2) + i\epsilon} \nonumber \\[0.2cm]
& & \label{eq:A3l} \\
A_4\!\!\!^{long} &=& \frac{e^5}{(2\pi)^4}  \langle Q',\lambda'| \, \gamma_{\mu} \, |P',\lambda' \rangle G_{\lambda,\lambda\!'}(t') \times \nonumber \\
& &\!\!\!\!\!\!\!\!\!\!\!\!\!\!\!\!\!\!\!\! \int \frac{d^4l  \langle Q,\lambda| \, \gamma_{\nu}(\not l +\rlap/ \!Q) \, \rlap/ \!K \, (\not l + \rlap/ \!Q + \rlap/ \!K)  \gamma^{\mu} (\not l + \rlap/ \!P) \gamma^{\nu} \, |P,\lambda \rangle}{(l^2 - {m_0}^2)((l+P)^2 - {m_e
}^2)((l+Q+K)^2 - {m_e}^2)((l+Q)^2 - {m_e}^2) + i\epsilon} \nonumber \\[0.2cm]
& & \label{eq:A4l} 
\end{eqnarray}

\noindent Using the identities

\begin{eqnarray}
\!\!\!\!\!\!\!\!\!(\not l + \rlap/ \!P - \rlap/ \!K) \, \rlap/ \!K\, (\not l + \rlap/ \!P) \!\!\! &=& \!\!\! (l+P)^2(\not l + \rlap/ \!P - \rlap/ \!K) - (l+P-K)^2(\not l+ \rlap/ \!P) \label{eq:idl1} \\ \!\!\!\!\!\!\!\!\!(\not l +\rlap/ \!Q)\, \rlap/ \!K \, (\not l + \rlap/ \!Q + \rlap/ \!K) \!\!\! &=& \!\!\!\! -(l+Q)^2(\not l + \rlap/ \!Q + \rlap/ \!K) + (l+Q+K)^2(\not l +\rlap/ \!Q) \label{eq:idl2}
\end{eqnarray}

\vspace{0.4cm}
\noindent one gets

\begin{eqnarray}
A_1\!\!\!^{long} &=& \frac{2e^5}{(2\pi)^4} \langle Q',\lambda'| \, \gamma_{\mu} \, |P',\lambda' \rangle G_{\lambda,\lambda\!'}(t') \times \nonumber \\
& & \int \mu^{\epsilon}d^nl \frac{\langle Q,\lambda| \, (\not l + \rlap/ \!P - \rlap/ \!K)\gamma^{\mu} (\not l + \rlap/ \!Q)              
 \, |P,\lambda \rangle}
  {(l^2 - {m_0}^2)((l+P-K)^2 - {m_e}^2)((l+Q)^2 - {m_e}^2) + i\epsilon} \nonumber  \\
& & \label{eq:A1lg} \\
A_2\!\!\!^{long} &=& - \frac{2e^5}{(2\pi)^4}  \langle Q',\lambda'| \, \gamma_{\mu} \, |P',\lambda' \rangle G_{
\lambda,\lambda\!'}(t') \times \nonumber \\
& & \int \mu^{\epsilon}d^nl \frac{ \langle Q,\lambda| \, (\not l + \rlap/ \!P) \gamma^{\mu} (\not l + \rlap/ \!Q + \rlap/ \!K) \, |P,\lambda \rangle}{(l^2 - {m_0}^2)((l+P)^2 - {m_e}^2)((l+Q+K)^2 - {m_e}^2) + i\epsilon} \nonumber  \\
& & \label{eq:A2lg} \\
A_3\!\!\!^{long} &=& - \frac{2e^5}{(2\pi)^4} \langle Q',\lambda'| \, \gamma_{\mu} \, |P',\lambda' \rangle G_{\lambda,\lambda\!'}(t') \times \nonumber \\
& & \left \{ \int \mu^{\epsilon}d^nl \frac{\langle Q,\lambda| \, (\not l + \rlap/ \!P - \rlap/ \!K)\gamma^{\mu} (\not l + \rlap/ \!Q)              
 \, |P,\lambda \rangle}
  {(l^2 - {m_0}^2)((l+P-K)^2 - {m_e}^2)((l+Q)^2 - {m_e}^2) + i\epsilon} \right. \nonumber  \\[0.2cm]
& & \left.- \int \mu^{\epsilon}d^nl \frac{ \langle Q,\lambda| \, (\not l + \rlap/ \!P) \gamma^{\mu} (\not l + \rlap/ \!Q ) \, |P,\lambda \rangle}{(l^2 - {m_0}^2)((l+P)^2 - {m_e}^2)((l+Q)^2 - {m_e}^2) + i\epsilon} \right \} \nonumber  \\
& & \label{eq:A3lg} \\
A_4\!\!\!^{long} &=& - \frac{2e^5}{(2\pi)^4} \langle Q',\lambda'| \, 
\gamma_{\mu} \, |P',\lambda' \rangle G_{\lambda,\lambda\!'}(t') \times 
\nonumber \\[0.2cm]
& & \left\{ \int \mu^{\epsilon}d^nl \frac{ \langle Q,\lambda| \, (\not l + \rlap/ \!P) \gamma^{\mu} (\not l + \rlap/ \!Q ) \, |P,\lambda \rangle}{(l^2 - {m_0}^2)((l+P)^2 - {m_e}^2)((l+Q)^2 - {m_e}^2) + i\epsilon} \right. \nonumber \\[0.2cm]
& & \left. - \int \mu^{\epsilon}d^nl \frac{ \langle Q,\lambda| \, (\not l + \rlap/ \!P) \gamma^{\mu} (\not l + \rlap/ \!Q + \rlap/ \!K) \, |P,\lambda \rangle}{(l^2 - {m_0}^2)((l+P)^2 - {m_e}^2)((l+Q+K)^2 - {m_e}^2) + i\epsilon} \right \} \nonumber  \\
& & \label{eq:A4lg} 
\end{eqnarray}

\noindent Thus it can readily be seen that 

\begin{equation}
A_1\!\!\!^{long}+A_2\!\!\!^{long}+A_3\!\!\!^{long}+A_4\!\!\!^{long} \; = \;0 
\;,
\end{equation}

\noindent which proves that $\{A_1+A_2+A_3+A_4 \}$ is gauge invariant. The proof for the other sums of amplitudes mentioned above proceeds analogously.

\section{The Internal Emission Gauge Variation} \label{sec:intgaugevar}

\noindent It follows from the definition of the vertex function \ref{eq:vertexf} and the decomposition \ref{eq:Gdecomp} that  

\begin{eqnarray}
A_3\!\!\!^{long} \!\!\! &=&  \frac{ie^5}{16\pi^2} \langle Q',\lambda'| \, \gamma_{\mu} \, |P',\lambda' \rangle G_{\lambda,\lambda\!'}(t') \times \nonumber \\
\!\!\!& & \langle Q,\lambda| \, \left\{ \Gamma^{\mu}(t',m_e\!^2-\alpha,m_e\!^2) \; - \; \Gamma^{\mu}(t,m_e\!^2,m_e\!^2 ) \right\} \, |P,\lambda \rangle \label{eq:wataka} \\
\!\!\!&=& \frac{ie^5}{16\pi^2} \langle Q',\lambda'| \, \gamma_{\mu} \, |P',\lambda' \rangle G_{\lambda,\lambda\!'}(t')
\langle Q,\lambda| \, \gamma^{\mu} \, |P,\lambda \rangle \left\{ F\!F \; - \; F\!F^0 \right\}  \;\;\; -  \nonumber \\
\!\!\!& & \frac{ie^5}{16\pi^2} \langle Q',\lambda'| \; \rlap/ \!Q \, |P',\lambda' \rangle G_{\lambda,\lambda\!'}(t')  
\langle Q,\lambda| \; \rlap/ \!K \, |P,\lambda \rangle \left\{ F\!F_a \; + \; F\!F_b \right\} \nonumber \\
\!\!\!&=& \frac{ie^5}{16\pi^2} \langle Q',\lambda'| \, \gamma_{\mu} \, |P',\lambda' \rangle G_{\lambda,\lambda\!'}(t') \langle Q,\lambda| \, \gamma^{\mu} \, |P,\lambda \rangle \times \nonumber \\ 
\!\!\!& & \left\{-4 + 2 B_0(m_e\!^2 - \alpha,m_0,m_e) - 2 B_0(m_0\!^2,m_e,m_e) + 3B_0(t,m_e,m_e) \right. \nonumber \\ \!\!\!& & -3B_0(t',m_e,m_e) - \frac{3\alpha}{t'+\alpha} (B_0(m_e\!^2 - \alpha,m_0,m_e) - B_0(t',m_e,m_e)) + \nonumber \\ \!\!\!& &  2tC_0(m_e\!^2,2m_e\!^2+t,m_e\!^2,m_0,m_e,m_e) - \nonumber \\ \!\!\!& & 2t'C_0(m_e\!^2-\alpha,2m_e\!^2+t',m_e\!^2,m_0,m_e,m_e) \left. \right\} \;\;- \nonumber \\
\!\!\!& & \frac{ie^5}{16\pi^2} \langle Q',\lambda'| \; \rlap/ \!Q \, |P',\lambda' \rangle G_{\lambda,\lambda\!'}(t')\langle Q,\lambda| \; \rlap/ \!K \, |P,\lambda \rangle \times \nonumber \\
\!\!\!& & \left\{ \right. -\frac{4t'}{t'+\alpha} C_0(m_e\!^2-\alpha,2m_e\!^2+t',m_e\!^2,m_0,m_e,m_e) + \nonumber \\
\!\!\!& & \frac{6t'-4\alpha}{(t'+\alpha)^2} B_0(m_e\!^2 - \alpha,m_0,m_e) - \frac{10t'}{(t'+\alpha)^2}B_0(t',m_e,m_e) \nonumber \\
\!\!\!& & + \frac{4}{t'+\alpha} \; (\frac{3}{2} \; + \; B_0(m_0\!^2,m_e,m_e)) \left. \right\} \label{eq:ffA3l}
\end{eqnarray}

\noindent The scalar integrals were written in explicit form in \ref{eq:ffA3l} since the FORM notation differs between the form-factor and internal emission routines. It is pivotal for the consistency check that Eq. \ref{eq:ffA3l} be compared to the FORM output after replacing $\rlap/ \varepsilon \rightarrow \rlap/ \!K $, which was done in the a3.gauge routine. The result reads, Ref. \cite{thesis}:

\begin{eqnarray}
A_3\!\!\!^{long} & = & \frac{ie^5}{16\pi^2} \langle Q',\lambda'| \, \gamma_{\mu} \, |P',\lambda' \rangle G_{\lambda,\lambda\!'}(t') \langle Q,\lambda| \, \gamma^{\mu} \, |P,\lambda \rangle \times \nonumber \\
& & \left\{ 2t C_{124} - 2t' C_{134}\!\!\!\!\!^{\alpha} - 2 B_{12} + 3 B_{24} + \frac{2t'-\alpha}{t'+\alpha} B_{13}\!\!\!^{\alpha} - \frac{3t'}{t'+\alpha} B_{34} \right\} \; + \nonumber \\
& &  \frac{ie^5}{16\pi^2} \langle Q',\lambda'| \; \rlap/ \!Q \, |P',\lambda' \rangle G_{\lambda,\lambda\!'}(t')\langle Q,\lambda| \; \rlap/ \!K \, |P,\lambda \rangle \times \nonumber \\
& & \left\{  \frac{4t'}{t'+\alpha} C_{134}\!\!\!\!\!^{\alpha} + \frac{2}{t'+\alpha} - \frac{4}{t'+\alpha} B_{12} + \right. \nonumber \\ & & \left. \frac{10t'}{(t'+\alpha)^2} B_{34} + \frac{4\alpha-6t'}{(t'+\alpha)^2} B_{13}\!\!\!^{\alpha}  \right\} \label{eq:invirA3l}
\end{eqnarray}

\noindent Eq. \ref{eq:invirA3l} is expressed with the notation of scalar integrals given in section \ref{sec:intemscalint} and can be seen to be identical to the result in \ref{eq:ffA3l} by observing that $B_0(m_0\!^2,m_e,m_e) = B_0(m_e\!^2,m_0,m_e) - 2 $. This therefore constitutes the strongest confirmation about the internal consistency of the calculation to this point since both results were derived in a completely independent manner!

\noindent The relation described by Eq. \ref{eq:wataka} is actually an example
of a higher order Ward-Takahashi identity \cite{itzzub,yfs61}. This can be seen by 
following Feynman's treatment of gauge invariance \cite{feyn2}, which leads to the 
following matrix identity \cite{yfs61} for the emission of a longitudinal photon
from an internal part of the electron line with four-momentum $P$ and possible
dependence on other external momenta $q_i$:

\begin{equation}
K_{\mu} \Lambda^{\mu}(P,K,q_i) \;\; = \;\; \Gamma(P-K,q_i) \; - \; 
\Gamma(P,q_i) \label{eq:mataka} \\
\end{equation}

\noindent The $\Gamma$-factors for non-zero transfer 
are directly related to self-energy
expressions and can be obtained from graphs \ref{eq:A3}, \ref{eq:A1}
and \ref{eq:A9} by removing the virtual photon connecting the two fermion 
lines. Thus, Eqs. \ref{eq:wataka} and \ref{eq:mataka} are two equivalent
representations of the Ward-Takahashi identity. The Ward-identity for
zero transfer will now be discussed in the following section.

\section{Ward Identity} \label{sec:Wardid}

\noindent Symmetries play important roles in physical theories not only because they allow for an elegant and insightful mathematical formulation of physics but also because they impose stringent conditions on the possible states of a system or on relations between certain quantities of the theory. One important consequence of the gauge symmetry in QED is the Ward identity \cite{bjdrell1}. Since it links self-energy and vertex functions in the limit of zero momentum
transfer, the calculated expressions in the respective chapters should be related according to: 

\begin{equation}
\Gamma^{\mu}(0,P^2,P^2) \;\;=\;\;- \; \frac{\partial \Sigma(P,m_0,m_e)}{\partial P_{\mu}} \label{eq:wardid}
\end{equation}

\noindent The l.h.s. of Eq. \ref{eq:wardid} can be derived by using the full result of Eq. \ref{eq:numvert} with both fermion legs off-shell and for zero transfer. This was done in the FORM routine $f\!fward$ in the appendix Ref. \cite{thesis}, and gives:

\begin{eqnarray}
\Gamma^{\mu}(0,P^2,P^2) &=& (B_0(P^2,m_0,m_e)-B_0(m_0\!^2,m_e,m_e)-4m_e\!^2C_{1230} -5) \gamma^{\mu} \nonumber \\ 
&&- \frac{2}{P^2}P^{\mu} \rlap/ \!P  \label{eq:wardlhs}
\end{eqnarray}

\noindent In order to obtain the r.h.s. of Eq. \ref{eq:wardid} one has to use the result for $\Sigma(P,m_0,m_e)$ in \ref{eq:sigma}. This leads to

\begin{eqnarray}
{- \; \frac{\partial \Sigma(P,m_0,m_e)}{\partial P_{\mu}}} \!\!\!&=&\!\!\! {- (4m_e\!^2C_{1230} +B_0(m_e\!^2,m_0,m_e) -B_0(P^2,m_0,m_e)+3) \gamma^{\mu}} \nonumber \\
&& + \; \frac{\partial B_0(P^2,m_0,m_e)}{\partial P_{\mu}} \; \rlap/ \!P \label{eq:wardrhsdef}
\end{eqnarray}

\noindent With

\begin{eqnarray}
\!\!\!\!\!\!\!&& \frac{\partial B_0(P^2,m_0,m_e)}{\partial P_{\mu}} \nonumber \\
\!\!\!\!\!\!\!&=&\frac{1}{i\pi^2} \int d^4l \frac{\partial}{\partial P_{\mu}} \frac{1}{(l^2-m_0\!^2)((l+P)^2-m_e\!^2)+ i \epsilon} \nonumber \\
\!\!\!\!\!\!\!&=& - \; \frac{2}{i\pi^2} \int d^4l \; \frac{l^{\mu}+P^{\mu}}{(l^2-m_0\!^2)((l+P)^2-m_e\!^2)^2 + i \epsilon} \nonumber \\
\!\!\!\!\!\!\!&=& - \frac{4}{i\pi^2} \int \! d^4l \int_0^1 \!\!dx \frac{x(l^{\mu}+P^{\mu})}{(l^2 + 2lPx - m_0\!^2 (1-x) + (P^2-m_e\!^2)x + i \epsilon)^3} \label{eq:dBfeyn}
\end{eqnarray}

\noindent and using \cite{pokorski}

\begin{eqnarray}
I_0 &=& \int d^nl \frac{1}{(l^2+2lP + M^2 + i\epsilon)^{\lambda}} = i(-\pi)^{\frac{n}{2}} \frac{\Gamma(\lambda-\frac{n}{2})}{\Gamma(\lambda)} \frac{1}{(M^2-P^2+ i\epsilon)^{\lambda-\frac{n}{2}}} \label{eq:feynint0} \nonumber \\
I_1 &=& \int d^nl \frac{l^{\mu}}{(l^2+2lP + M^2 + i\epsilon)^{\lambda}} = -P^{\mu} \; I_0 \label{eq:feynint1}
\end{eqnarray}

\noindent one finally gets

\begin{eqnarray}
\frac{\partial B_0(P^2,m_0,m_e)}{\partial P_{\mu}} &=& 2 P^{\mu} \int_0^1 dx \frac{-x^2+x}{P^2x^2-(P^2-m_e\!^2+m_0\!^2)x + m_0\!^2 + i\epsilon} \nonumber \\
&=& \frac{2 P^{\mu}}{P^2} \int_0^1 dx \left\{-1 + \frac{\frac{m_e\!^2}{P^2}x}{x^2-\frac{P^2-m_e\!^2+m_0\!^2}{P^2}x+\frac{m_0\!^2}{P^2} + i \epsilon} \right\} \nonumber \\
&=& - \frac{2 P^{\mu}}{P^2} \label{eq:dBres}
\end{eqnarray}

\noindent up to ${\cal O}(\frac{m_e\!^2}{P^2})$. Inserting \ref{eq:dBres} into \ref{eq:wardrhsdef} and using the relation

\begin{equation}
B_0(m_e\!^2,m_0,m_e) \; = \; 2 \; + \; B_0(m_0\!^2,m_e,m_e) \label{eq:Brel}\;,
\end{equation}

\noindent which was already used in section \ref{sec:intgaugevar}, yields the identical result to Eq. \ref{eq:wardlhs}. This proves that the Ward identity \ref{eq:wardid} is satisfied by the expressions for the self-energy and vertex functions, calculated in this work in completely independent ways!

\chapter{Monte Carlo Results} \label{sec:MCres}

\section{Finite Mass Effects} \label{sec:masseff}

\noindent In the previous chapters only massless fermions were considered. For high energy processes, this is a very good approximation when $ \frac{m^2}{-t} \ll 1$ as long as the radiated photon is not emitted in a direction parallel to one of the fermion directions. Within the context of the theory of multiple bremsstrahlung in gauge theories at high energies \cite{calkul1}, a method was 
developed to incorporate finite mass effects into a description using 
massless fermion spinors \cite{calkul3}. This is especially necessary when running 
Monte Carlo simulations, where a bulk of events is generated in the low angle emission area of phase space and a wrong collinear limit would render numerical results meaningless. Following the CALKUL collaboration \cite{calkul2}, the mass terms important in the collinear limit have the following general form in the single bremsstrahlung case:

\begin{equation}
\frac{d\sigma^m}{K^0 dK^0 d\Omega_K d \Omega_{e^+}} \; = \; \frac{(Q'^0)^2}{2^9 \pi^5 s s'}
\; |A^m|^2 \; , \label{eq:csmasscorr}
\end{equation}

\noindent with

\begin{equation}
|A^m|^2 \;\;=\;\; - \; \frac{e^2m^2}{(qk)^2} f^0(q-k,p_i) \;, \label{eq:genmasscorr}
\end{equation}

\noindent where the photon is radiated nearly parallel to $q$, and $f^0$ denotes the non radiative cross section, summed over all polarizations, with the original $q$ replaced by $q-k$. In the case of Bhabha scattering the Born cross section is proportional to the following invariant summed matrix element squared \cite{calkul2}: 

\begin{equation}
f^B\!\!\!\!_{e^-+e^+} \;\;=\;\; \frac{2e^4}{t^2}(s^2+u^2)
\end{equation}

\noindent The complete non-radiative cross section for the ${\cal O}(\alpha^2)$ single bremsstrahlung mass corrections is then proportional to

\begin{equation}
f^0\!\!\!_{e^-+e^+} \;\;=\;\;(1+ \frac{e^2}{4\pi^2} 
f\!f^0)f^B\!\!\!\!_{e^-+e^+}
\end{equation}

\noindent with 

\begin{eqnarray}
f\!f^0(t) &\equiv& F\!F^0(t) - 4\pi Re B(t) \nonumber \\
&=& -\; 2 + 2 ln \frac{-t}{m_e\!^2} \label{eq:softff}
\end{eqnarray}

\noindent From Eq. \ref{eq:genmasscorr} it follows that, when summed over all fermion legs, the finite mass terms for the ${\cal O}(\alpha^2)$ single bremsstrahlung corrections are given by

\begin{eqnarray}
|A^{m_e}\!\!\!\!_{e^- + e^+}|^2 &=&  - \; \frac{2e^6m_e\!^2}{(PK)^2} \left[ 1 + \frac{e^2}{4\pi^2} f\!f^0 (-2PQ+2QK) \right] \;\; \times \nonumber \\ &&  \frac{(2PP'-2P'K)^2 + (-2PQ'+2Q'K)^2}{(-2PQ+2QK)^2} \nonumber \\
&& - \; \frac{2e^6m_e\!^2}{(QK)^2} \left[ 1 + \frac{e^2}{4\pi^2} f\!f^0 (-2PQ-2PK) \right] \;\; \times \nonumber \\ && \frac{(2QQ'+2Q'K)^2 + (-2QP'-2P'K)^2}{(-2PQ-2PK)^2} \nonumber \\
&& - \; \frac{2e^6m_e\!^2}{(P'K)^2} \left[ 1 + \frac{e^2}{4\pi^2} f\!f^0 (-2P'Q'+2Q'K) \right] \;\; \times \nonumber \\ && \frac{(2PP'-2PK)^2 + (-2QP'+2QK)^2}{(-2P'Q'+2Q'K)^2} \nonumber \\
&& - \; \frac{2e^6m_e\!^2}{(Q'K)^2} \left[ 1 + \frac{e^2}{4\pi^2} f\!f^0 (-2P'Q'-2P'K) \right] \;\; \times \nonumber \\ && \frac{(2QQ'+2QK)^2 + (-2PQ'-2PK)^2}{(-2P'Q'-2P'K)^2} 
\end{eqnarray}

\section{The Soft Limit} \label{sec:softl}

\noindent The soft limit of the exact ${\cal O}(\alpha^2)$ expression is determined
 by combining all helicity-independent($hi$) terms of the result \ref{eq:exres} multiplying the lower order amplitude. Doing this and using Eq. \ref{eq:C124res}
and Ref. \cite{holl86} gives the following:

\begin{eqnarray}
A^{hi}_{e^-} &=& \frac{e^2}{16\pi^2} \left\{ -8 -8m_e\!^2C^0_{123} - 2(t C_{124} + t'
C_{134}\!\!\!\!\!\!^0 \;\;) + 6 B_{12} - 6
   B_{34} \right\} A^{TL}\!\!\!\!\!\!_{e^{-}} \nonumber \\
&=& \frac{e^2}{16\pi^2} \left\{ - 8 + \frac{2\pi^2}{3} - ln^2
\frac{-t}{m_e\!^2} - ln^2
  \frac{-t'}{m_e\!^2} + 6 ln \frac{-t'}{m_e\!^2} \right. \nonumber \\
& & \left. + 4 ln \frac{-t}{m_e\!^2} ln \frac{m_0}{m_e} + 4 ln 
\frac{-t'}{m_e\!^2
} ln \frac{m_0}{m_e} - 8 ln \frac{m_0}{m_e}\right\} A^{TL}\!\!\!\!\!\!_{e^{-}} \label{eq:hi}
\end{eqnarray}

\noindent Using the virtual YFS-IR function from Eq. \ref{eq:Btres}
 one gets

\begin{eqnarray}
&& A^{hi}_{e^-} - \alpha \left[ Re \; B(t) + Re \; B(t') \right] A^{TL}\!\!\!\!\!\!_{e^{-}} \nonumber \\ &=& \frac{e^2}{16\pi^2} \left\{ - 4 + 5 ln \frac{-t'}{m_e\!^2} - ln \frac{-t}{m_e\!^2} \right\} A^{TL}\!\!\!\!\!\!_{e^{-}} \label{eq:beta1} 
\end{eqnarray}

\noindent In the soft limit $t=t'$ so that 

\begin{eqnarray}
A^{sl}_{e^-} - 2\alpha Re \; B(t) A^{TL}\!\!\!\!\!\!_{e^{-}} 
&=& \frac{e^2}{4\pi^2} \left\{ - 1 + ln \frac{-t}{m_e\!^2} \right\} A^{TL}\!\!\!\!\!\!_{e^{-}} \nonumber \\ 
&=& \frac{e^2}{8\pi^2} \; f\!f^0(t) \; A^{TL}\!\!\!\!\!\!_{e^{-}} \;, \label{eq:sl}
\end{eqnarray}

\noindent where $f\!f^0$ is given by Eq. \ref{eq:softff}. Taking into 
account that the cross section is proportional to 
$2 Re\{A^{TL}\!\!\!\!\!\!_{e^{-}} * A_{e^-}\!\!\!\!\!^* \; \}$, the soft limit
contains the identical form factor as the mass correction term in 
\ref{eq:softff}. This is another consequence from the YFS-theory discussed
in chapter \ref{sec:YFS} which states that in the soft limit, the summed
and averaged bremsstrahlung cross section factorizes into a soft part given
by $\widetilde{S}(k)$, Eq. \ref{eq:YFSsdef}, and
the lower order cross section, theorem \ref{eq:YFSbeta}.

\noindent It is useful in Eq. \ref{eq:beta1} to keep $t$ and
$t'$ separately, since the formula 
holds also for large values of
$k=\frac{E_{\gamma}}{E_{Beam}}$ as numerical results in Fig. \ref{fig:NLL1} show.
In order to compare Eq. \ref{eq:hi} with the result obtained by Fadin
et al. \cite{kuraev94}, one has to use the real YFS-IR function given in \ref{eq:Bttildedef}.
This then leads to

\begin{eqnarray}
&& A^{hi}_{e^-} + \alpha \left[ \widetilde{B}(t,E_{cut}) +  \widetilde{B}(t',E_{cut}) \right] A^{TL}\!\!\!\!\!\!_{e^{-}} \nonumber \\ 
&=& \frac{e^2}{8\pi^2} \left\{ - 4
+ln \frac{-t}{{m_e}\!^2}(2ln \Delta - ln x ) \right.
\nonumber
 \\ & & \left. + ln \frac{-t'}{{m_e}\!^2} (3 + 2 ln \Delta )  - 4 ln \Delta +  ln x \right\} A^{TL}\!\!\!\!\!\!_{e^{-}}   \label{eq:comprus}
\end{eqnarray}

\noindent with $x=\frac{E_Q}{E_P}$ and $\Delta= \frac{E_{cut}}{E_P}$ (in 
the cm-system $E_P=E_{Beam}$). Both Eq. \ref{eq:beta1} and 
\ref{eq:comprus} are
 normalized such that they multiply $\frac{\alpha}{\pi} d\sigma ^{{\cal O}(\alpha)}$
as is the case in Ref. \cite{kuraev94}. The
difference to the formula given by Fadin et al. is $\frac{1}{4} 
+ \frac{1}{2} ln^2 x$. 
All equations are valid for electron-line emission. The other
 line must be added analogously.

\section{Leading Log Comparisons} \label{sec:llog}

\noindent At this point all the pieces needed for a Monte Carlo 
integration are calculated and checked on a differential level. It is now
necessary to combine the various results and include the phase space factors
in order to get a normalized differential cross section. Following the 
conventions of Ref. \cite{bjdrell1,bjdrell2}, the expression for 
$d\sigma^{{\cal O}(\alpha+\alpha^2)}$ reads:

\begin{eqnarray}
\frac{d\sigma^{{\cal O}(\alpha+\alpha^2)}}{K^0 dK^0 d\Omega_K d\Omega_{e^+}} 
\!\!\! & = & \!\!\! 
\frac{(Q'^0)^2}{2^{9} \pi^5 s s'}  \sum_{\rho,\lambda,\lambda'} 
\; Re \left\{ \left( 
A^{TL}\!\!\!\!\!\!_{e^{-}} \; + \; A^{TL}\!\!\!\!\!\!_{e^{+}} \;\;
\right) \left[
A_{e^-}\!\!\!\!\!^* \;\; + \; A_{e^+}\!\!\!\!\!^* \; \right] \right\} \;\;+
\nonumber \\ \!\!\! && \!\!\! \frac{(Q'^0)^2}{2^{10} \pi^5 s s'}  
\sum_{\rho,\lambda,\lambda'}\Big\vert A^{TL}\!\!\!\!\!\!_{e^{-}}  +  
A^{TL}\!\!\!\!\!\!_{e^{+}} \Big\vert^2  + 
\frac{d\sigma^m}{K^0 dK^0 d\Omega_K d\Omega_{e^+}} 
\label{eq:normcs}
\end{eqnarray}

\noindent The amplitudes $A_e$ are expressed in Eq. \ref{eq:exres}; it is 
assumed that all helicity degrees of freedom have been summed over 
and that the 
initial states were averaged. This cross section still contains the IR-mass 
divergent terms
proportional to $ln \frac{m_0}{m_e}$, which are identical to those given in
Eq. \ref{eq:hi}. 
Following the YFS-prescription of chapter \ref{sec:YFS}, the normalized
infrared regulated (reg)
differential cross section for the exact result \ref{eq:exres}
is then given by:

\begin{eqnarray}
\frac{d{\sigma^{{\cal O}(\alpha+\alpha^2)}\!\!\!\!\!\!\!\!\!\!\!\!\!\!\!\!\!\!\!\!_{reg}}}{K^0 dK^0 d\Omega_K d\Omega_{e^+}}
&=& \frac{d\sigma^{{\cal O}(\alpha+\alpha^2)}}{K^0 dK^0 d\Omega_K d\Omega_{e^+}} \; - \nonumber \\
&& 2 \alpha Re \left\{ B(t) + B(t') \right\}  
\frac{d\sigma^{{\cal O}(\alpha)}}{K^0 dK^0 d\Omega_K d\Omega_{e^+}} \label{eq:excs}
\end{eqnarray}

\noindent An analytic five dimensional integration over \ref{eq:excs} is
extremely involved and beyond known calculational techniques. It is also 
desirable to be able to change to a given detector geometry and new
experimental cutoff parameters without having to do a completely new calculation. Both of these difficulties can be overcome by using a Monte Carlo (MC) 
approach \cite{sobol}. The algorithm implemented uses the method of  
weighting \cite{sobol}. Events, i.e. sets of final state four-momenta, 
are generated at random, with a probability for each configuration to occur
given by \ref{eq:excs}. Any experimental situation can then easily be 
simulated by throwing away those events that do not satisfy the experimental
conditions. The generation of events can be accomplished by first generating
"trial" events according to some approximate cross section $\frac{d \sigma_{approx}}{K^0 dK^0 d\Omega_K d\Omega_{e^+}}$, and assigning to each trial event a weight

\begin{equation}
w \; = \; \left(\frac{d{\sigma^{{\cal O}(\alpha+\alpha^2)}\!\!\!\!\!\!\!\!\!\!\!\!\!\!\!\!\!\!\!\!_{reg}}}{K^0 dK^0 d\Omega_K d\Omega_{e^+}} \right) \; \left( \frac{d \sigma_{approx}}{K^0 dK^0 d\Omega_K d\Omega_{e^+}} \right)^{-1} \label{eq:weight}
\end{equation}

\noindent The exact event distribution is then realized by accepting the trial
events with a probability proportional to $w$. From the above considerations
it is clear that $\frac{d \sigma_{approx}}{K^0 dK^0 d\Omega_K d\Omega_{e^+}}$ must be as 
simple as possible and that the method is only practical if the weights 
\ref{eq:weight} do not fluctuate too wildly. The approximate cross section
chosen here is the one suggested by F.A.Berends and R.Kleiss \cite{berends83}
and was implemented into Fortran by S.A.Yost. In order to be able to generate
the phase space variables $K^0, \theta_K, \Phi_K, \theta_{e^+}$ and 
$\Phi_{e^+} \;$, the integral over $d \sigma_{approx}$ has to be 
known and some
cuts for these parameters have to be chosen. Here, the algorithm of Ref. 
\cite{berends83} was used and the cuts correspond to the SICAL W-N acceptance
\cite{Been93}. 

\noindent The only practical way to calculate $\int w d \sigma_{approx}$ is
to generate uniform distributions $dq_i$ between $[0,1]$ for each differential
factor and write:

\begin{equation}
d \sigma_{approx} \;=\; J dq_{K^0} dq_{\theta_K} dq_{\Phi_K}
dq_{\theta_{e^+}} dq_{\Phi_{e^+}} \;, \label{eq:dapprox}
\end{equation}

\noindent where $J$ is the Jacobian of the transformation \ref{eq:dapprox}.
The integrated cross section is then given by summation of all accepted
weights and averaging over the total number of generated events.

\begin{center}
\begin{figure}
\centering
 \epsfig{file=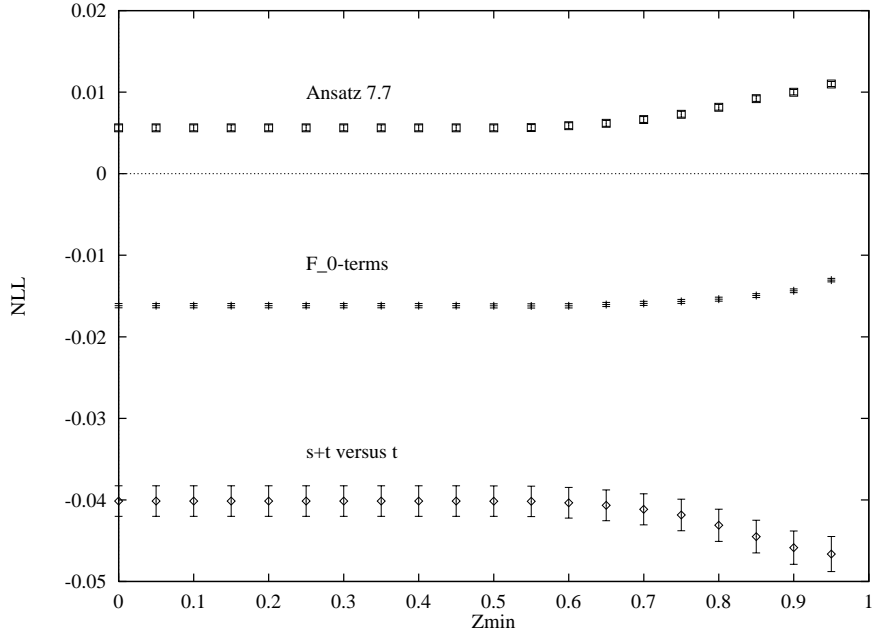,height=12cm,angle=270}
 \caption{Results of a Monte Carlo integration corresponding to $10^5$
          events per point; $NLL \equiv 
          \{{\sigma^i\!\!_{reg}} \; - {\sigma^{{\cal O}(\alpha^2)}\!\!\!\!\!\!\!\!\!\!\!\!\!\!_{reg}} \;\;\;\,\}
         $/${\sigma^{{\cal O}(\alpha^2)}\!\!\!\!\!\!\!\!\!
         \!\!\!\!\!_{reg}} \;\;\;$ is a measure for the 
         size of the subleading and pure $\left(\frac{\alpha}{\pi}\right)^2$
          terms in the various cases (i). The real photon IR-singularity is
         regulated by a cutoff and cancels out of the ratios. 
         The energy cut $z_{min}$ is defined
         in Ref. \protect\cite{JadRiWa95}, for example, as is the W-N acceptance.}
 \label{fig:NLL1}
\end{figure}
\end{center}

\vspace{-1cm}
\noindent The results for the integrated exact ${\cal O}(\alpha^2)$
cross section
are presented in Fig. \ref{fig:NLL1} in comparison
with several partial results following from the exact expression
\ref{eq:exres}. It can be seen that the ansatz \ref{eq:hi} stays within
$1\%$(!) of the exact t-channel expression \ref{eq:exres} for the chosen
cuts. This is a remarkable result given the simplicity of that equation but
should be seen as somewhat accidental, since many helicity dependent terms
multiplying the tree-level amplitude apparently cancel, without any obvious
physical necessity. Taking into account only ${\cal F}_0$ terms is also in
excellent agreement with the complete t-channel result and remains within
$2\%$. The effect of adding the s-channel, including $Z$-exchange, can be seen
to be relatively small, roughly $5\%$ of the second order single
bremsstrahlung expression \ref{eq:exres} for the chosen parameter space.
In conclusion, for the accuracy required in the context of $0.1\%$ precision
radiative corrections, both t-channel approximations \ref{eq:hi} and using 
only ${\cal F}_0$ terms in \ref{eq:exres} represent perfectly acceptable
approaches.

\begin{center}
\begin{figure}
\centering
 \epsfig{file=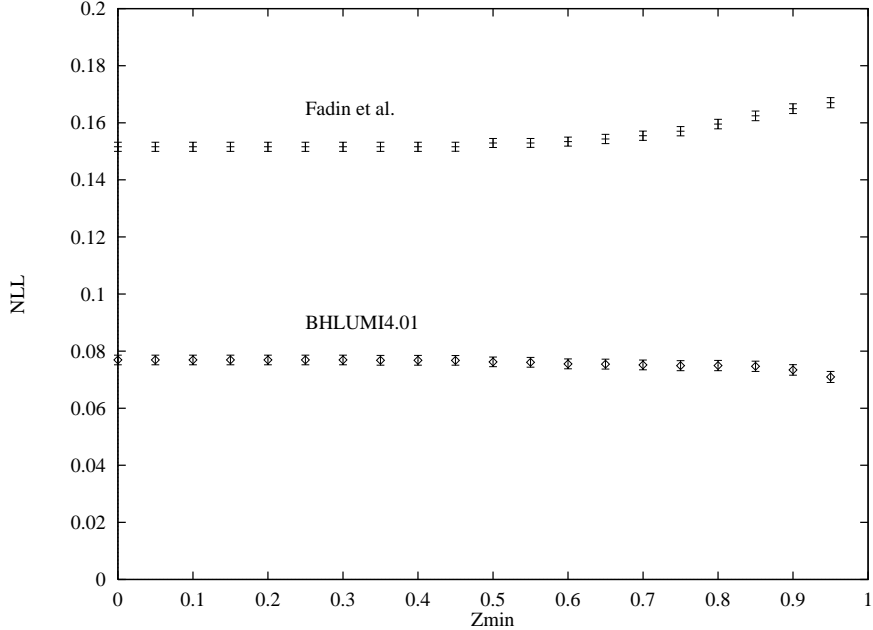,height=12cm,angle=270}
 \caption{A comparison of the integrated cross section with two independent
          calculations, analogous to Fig. \protect\ref{fig:NLL1},
          corresponding to $10^5$ events per point. The result
          \protect\ref{eq:exres}, including s and t channel plus $Z$-exchange
          and IR-regularization, is seen to deviate in its non leading 
          logarithmic parts from the pure t-channel expression given in Ref.
          \protect\cite{kuraev94} by up to 17$\%$. The LL-ansatz
          implemented in BHLUMI4.01 \protect\cite{JadRiWa95} stays within
          8$\%$ over the 
          entire range of $z_{min}$.} 
 \label{fig:NLL2}
\end{figure}
\end{center}

\vspace{-1cm}
\noindent As a further consistency check and also as a measure for the size
of the hitherto missing subleading terms from the second order single
bremsstrahlung calculation, the complete result derived in this work with
the s-channel and $Z$-exchange is compared in Fig. \ref{fig:NLL2} to two
independent leading log type results. The size of the overall subleading parts
is very much dependent of what exactly is included in the leading log
ansatz and the authors of Ref. \cite{kuraev94} claim to have all terms 
contributing to a $0.1\%$ precision level in their formula. While an exact
value for the size of the missing subleading bremsstrahlung parts is difficult
to specify, Fig. \ref{fig:NLL2} suggests that the total contribution 
of those terms is below $20\%$ of the total pure second order result.
Since the size of the complete second order contribution is roughly
$0.35\%$ of the Born process for the W-N acceptance at SICAL \cite{Been93}, 
the subleading terms are found to be  
small but non negligible. The good agreement of the LL-ansatz in BHLUMI4.01,
however, suggests that the accuracy of the theoretical precision level 
quoted in Refs. \cite{JadWas95,Gatlin94,Yellow95} as ${\cal O} (0.16\%)$
relative to the Born process, is rather conservative indeed.

\chapter{Conclusions} \label{sec:concl}

\noindent In this thesis, the sole outstanding contribution to low angle 
Bhabha scattering above the $0.1\%$ precision level, the second order single
bremsstrahlung subleading terms, have been calculated exactly, i.e. including
pure $\left(\frac{\alpha}{\pi} \right)^2$ corrections. While other leading log
and partial results have been published \cite{berends87,kuraev94}, Eq. \ref{eq:exres} represents
the first complete, fully differential exact ${\cal O}(\alpha)$ corrections to
$e^- \; + \; e^+\; \longrightarrow \;e^- \;+\; e^+\; + \; \gamma \;$ 
in the LEP/SLC energy
regime. The accuracy of the presented result is well below $0.05\%$, which is
the level required in order to have meaningful comparisons with measurements
of the new luminosity detectors at LEP \cite{Piet94}. The calculated 
expression \ref{eq:exres} has passed several strong internal consistency
checks as well as analytical and numerical comparisons with leading log type
results:

\vspace{0.3cm}
\noindent It was shown in chapter \ref{sec:YFS} that the expression 
for the internal emission 
amplitudes \ref{eq:a3a4res} is indeed UV-finite and possesses the expected
logarithmic virtual photon mass dependence as predicted by the Yennie,
Frautschi and Suura program. 

\noindent Chapter \ref{sec:gaugevar} provided an essential link
between technically completely unrelated parts of the calculation in deriving
the gauge variation for the initial state internal emission graph \ref{eq:A3}
by replacing $\varepsilon^{\mu} \rightarrow K^{\mu}$ and comparing this
with the analytical expressions for the on- and off-shell form factors.
The demonstrated agreement represents a Ward-Takahashi identity for non-zero
momentum transfer.
In addition to being a very strong check on the internal emission result,
it also demonstrates the correct application of the on-shell renormalization
scheme employed in this work.
In section \ref{sec:Wardid}, the other remaining part of the 
calculation, the self-energy contribution, was linked
to the off-shell form factor in
the limit of zero transfer. The verified Ward identity completes the full
circle of all the various terms given in this thesis and establishes 
powerful evidence for the overall correctness of the derived expressions.

\noindent In chapter \ref{sec:MCres} the mass corrections showing a double 
pole structure, the only relevant mass terms in the high energy approximation
for a Monte Carlo integration,
were derived and subsequently added to yield the complete result. 
The soft limit
of Eq. \ref{eq:exres} was given explicitly in differential form in
\ref{eq:sl} and shown
to be connected to the lower order cross section according to YFS-theory. 
By combining all helicity independent terms, an approximate formula
\ref{eq:hi} was found that shows agreement with the exact result within
$1\%$(!) after a Monte Carlo phase space integration for SICAL-type 
cuts. It was also shown that the terms proportional to the tree-level cross
section make up for roughly $98\%$ of the second order t-channel 
contribution and thus
contain all relevant leading and subleading logarithmic corrections. The size
of the s-channel corrections including $Z$-exchange was found to be as small
as expected, around $5\%$ relative to the pure t-channel result. 
Further MC results demonstrated that the exact expression \ref{eq:exres}
relates to other leading log type calculations as follows:

\vspace{0.3cm}
\noindent The IR$-$regulated ratios 
$\{{\sigma^i\!\!_{reg}} \;-{\sigma^{{\cal O}(\alpha^2)}\!\!\!\!\!\!\!\!\!\!\!\!\!\!_{reg}} 
\;\;\;\, \} $/${\sigma^{{\cal O}(\alpha^2)}\!\!\!\!\!\!\!\!\!\!\!\!\!\!_{reg}} \;\;\;$, which are a measure
of the size of the missing subleading terms in the various 
partial results ($i$),
revealed that the calculation by Fadin et al. lies within $17\%$ of the exact
formula \ref{eq:exres}.
It was furthermore shown that the ansatz used in BHLUMI4.01 \cite{JadWas95} 
is within $8\%$ of the
answer derived in this work for the specified cuts. Since the size of the
${\cal O}(\alpha^2)$ single bremsstrahlung contribution is of the order of
$0.35\%$ relative to the Born process for cuts chosen to correspond to the
experimental luminosity detector SICAL at ALEPH, 
this will have important consequences 
for the overall level of precision of BHLUMI4.x. While several questions about
the dependence on experimental cuts for the above agreement as well as those 
relating to the technical precision domain remain to be discussed more 
thoroughly \cite{Melles95,Melles96}, it is safe to assume that the precision of BHLUMI4.x will
be below the $1pm$ level.

\vspace{0.3cm}
\noindent In light of the results presented in this thesis, it will soon be
realistic to achieve a theoretical level of accuracy below $1pm$ for the
total cross section in low angle Bhabha scattering. Together with high 
precision measurements of second generation luminosity detectors at CERN,
it will hopefully be possible to determine various electroweak parameters 
in a way that sets stringent conditions on the range of validity of this  
theory and, simultaneously, to open up a new regime in the search for  
physics beyond the Standard Model.

\section*{Acknowledgments}

\noindent I would like to thank B.F.L. Ward for suggesting this project for
my doctoral thesis as well as his support throughout its duration. By the same
token I am obliged to S.A. Jost for the abovementioned contributions and 
likewise to S. Jadach. In addition, my gratitude extends to H.D. Dahmen for  
his support of my doctoral thesis from abroad. Finally, I am grateful to 
S.J. Brodsky and the SLAC-theory group for their kind hospitality.

\bibliographystyle{plain}
\bibliography{books,jadward,calkul,genthy,genexp,bstmod,melles}

\end{document}